\documentclass[aps,amsmath,amssymb,nofootinbib,preprintnumbers,twocolumn,showpacs]{revtex4}
\usepackage{epsfig}
\usepackage{bbm}
\usepackage{mathrsfs,graphicx,bm,amsmath}
\usepackage{hyperref}
\usepackage{color}

\definecolor{holger}{rgb}{0,0.4,0.7}
\definecolor{jan}{rgb}{0.7,0,0.4}
\definecolor{sebastian}{rgb}{0.7,0,0}
\definecolor{comment}{rgb}{0.9,0,0}
\newcommand{\kF}{k_{\text{F}}}

\newcommand{\epm}{\epsilon_{\text{M}}}

\newcommand{\cl}{\text{cl}}

\newcommand{\hpb}{\bar{h}_\varphi}

\def\di{\displaystyle}

\def\bg{\begin{eqnarray}\begin{array}{rcl}\displaystyle}
\def\eg{\end{array} &\di    &\di   \end{eqnarray}}
\def\bm#1{\begin{eqnarray}\begin{array}{#1}\di}
\def\bmo#1{\begin{eqnarray*}\begin{array}{#1}\di}
\def\bml#1#2{\begin{eqnarray}\begin{array}{#1}\label{#2}\di}
\def\bgo{\begin{eqnarray*}\begin{array}{rcl}\displaystyle}
\def\ego{\end{array} &\di    &\di \nonumber  \end{eqnarray*}}

\def\btensor#1#2{\renew\left#1\begin{array}{#2}\di}
\def\brtensor#1#2#3{\ren#3\left#1\begin{array}{#2}}
\def\botensor#1#2{\renew\left#1\begin{array}{#2}}
\def\etensor#1{\end{array}\right#1}

\def\eq#1{(\ref{#1})}
\def\Eq#1{Eq.~(\ref{#1})}

\def\s0#1#2{\mbox{\small{$ \frac{#1}{#2} $}}}
\def\0#1#2{\frac{#1}{#2}}

\begin{document}


\title{Functional renormalization group approach to the BCS-BEC crossover}

\author{S. Diehl${}^{a}$}
\author{S. Floerchinger${}^{b}$}
\author{H. Gies${}^{b,c}$}
\author{J. M. Pawlowski${}^{b}$}
\author{C. Wetterich${}^{b}$}

\affiliation{
\mbox{\it ${}^a$Institute for Quantum Optics and Quantum
Information of the Austrian Academy of Sciences,}\\
\mbox{\it A-6020 Innsbruck, Austria}\\
\mbox{\it ${}^b$Institut f{\"u}r Theoretische Physik,
Philosophenweg 16, D-69120 Heidelberg, Germany}\\
\mbox{\it ${}^c$Theoretisch-Physikalisches Institut, Friedrich-Schiller-Universit\"at,
Max-Wien-Platz 1, D-07743 Jena, Germany}\\}



\begin{abstract}
The phase transition to superfluidity and the BCS-BEC crossover for an
ultracold gas of fermionic atoms is discussed within a functional
renormalization group approach. Non-perturbative flow equations, based on an
exact renormalization group equation, describe the scale dependence of the
flowing or average action. They interpolate continuously from the microphysics
at atomic or molecular distance scales to the macroscopic physics at much
larger length scales, as given by the interparticle distance, the correlation
length, or the size of the experimental probe. We discuss the phase diagram as
a function of the scattering length and the temperature and compute the gap,
the correlation length and the scattering length for molecules. Close to the
critical temperature, we find the expected universal behavior. Our approach
allows for a description of the few-body physics (scattering and molecular
binding) and the many-body physics within the same formalism.
\end{abstract}


\maketitle


\section{Introduction}

The binding energy of molecules formed from two fermionic atoms may depend on
external parameters such as a magnetic field $B$. We discuss a gas of two
degenerate species of atoms (typically hyperfine states), where bound
molecules form for $B<B_0$, whereas for $B>B_0$ only unstable resonance states
are present. At the Feshbach resonance $B=B_0$, the scattering length $a$
diverges. At low temperatures $T$ and for large $B-B_0$, the molecules play no
role and one expects superfluidity of the BCS type \cite{BCS}. In the opposite
case of large negative $B-B_0$, the bosonic molecules dominate at low
temperatures $T$ and Bose-Einstein condensation (BEC) \cite{BoseEinstein}
should occur. In the vicinity of the Feshbach resonance, a continuous crossover
from BCS to BEC condensation has been predicted \cite{p5ALeggett80}. Several
recent experiments have established this overall picture \cite{Exp}. In
parallel, a considerable theoretical effort has been devoted to a
quantitatively precise theoretical understanding of this phenomenon
\cite{Nussinov, Nishida:2006br, Nishida:2006eu, EpsilonExpansion,
  Nikolic:2007zz, AbukiBrauner, MonteCarlo, MonteCarloBulgac, MonteCarloBuro,
  MonteCarloAkki, TMatrix, Diehl:2005ae, Diener08, Haussmann:2007zz, Birse05,
  Diehl:2007th, Diehl:2007XXX, Gubbels:2008zz, FSDW08,KopLo09}.

A major motivation for the theoretical effort is due to the possibility of
using this system as a testing ground for non-perturbative methods in
many-body theory, which go beyond the mean-field approximation. Quantitatively
reliable methods for complex many-body systems are crucial for many areas in
physics, from condensed matter phenomena to nuclear or elementary particle
physics. For a non-relativistic system, as in our case, the results of such
methods may be tested, in principle, by a huge number of experiments and
observations. However, for a quantitatively precise test the underlying
microphysics must be known with sufficient precision. This is often not the
case for real physical systems in condensed matter physics. A notable
exception is the universal physics near a second order phase transition, where
the critical exponents and amplitude ratios no longer depend on microscopic
details. The quantitative computations of these universal quantities by the
renormalization group has been one of the major breakthroughs in theoretical
physics \cite{Wilson, Wegner}.

In cold atom gases, the microscopic physics is well under control, since it can
be directly related to atomic and molecular properties and measured by
scattering experiments. This gives rise to the hope that experimental tests
become now available for the whole connection between microphysics on the one
and the macroscopic quantities as thermodynamics, condensation phenomena, the
correlation length, defects etc.\ on the other side. Furthermore, the
effective strength of interaction can be tuned experimentally, which offers
the fascinating perspective of following the tests of theoretical methods from
the well-controlled perturbative domain for small scattering length to the
difficult non-perturbative region for large scattering length.

First results for the BCS-BEC crossover are now available for a number of
theoretical methods that go beyond mean-field theory. Quantitative
understanding of the crossover at and near the resonance has been developed
through numerical calculations using various quantum Monte-Carlo (QMC) methods
\cite{MonteCarlo,MonteCarloBulgac,MonteCarloBuro,MonteCarloAkki}. Computations
of the complete phase diagram have been performed from functional
field-theoretical techniques, in particular from $\epsilon$-expansion
\cite{Nussinov, Nishida:2006br, Nishida:2006eu, EpsilonExpansion},
$1/N$-expansion \cite{Nikolic:2007zz, AbukiBrauner}, $t$-matrix approaches
\cite{TMatrix}, Dyson-Schwinger equations \cite{Diehl:2005ae, Diener08},
2-Particle Irreducible methods \cite{Haussmann:2007zz}, and
renormalization-group flow equations \cite{Birse05, Diehl:2007th,
  Diehl:2007XXX, Gubbels:2008zz, FSDW08,KopLo09}.

In this paper, we give an account of the functional renormalization group
approach for the BCS-BEC crossover which is based on the flowing action (or
average action) \cite{Wetterich:1992yh}. We concentrate here on the detailed
description of the formalism which underlies the results presented in
ref.\ \cite{Diehl:2007th, Diehl:2007XXX, FSDW08}. Furthermore, we extend the
results of these papers by results for the correlation length and the
self-interaction of the bosonic molecules, including the behavior near the
phase transition.

Our paper is organized as follows. Sect.\ \ref{Method} focuses on the method which we use in order to
tackle the crossover problem, the functional renormalization group. Our starting point is an exact functional renormalization group equation for the average potential \cite{Wetterich:1992yh}. We specify the approximation scheme used for our practical computations and present the non-perturbative flow equations. Details and various quantities needed for a numerical solution can be found in a series of appendices A-F. This section provides the technical basis for our
work. 

The physics of the crossover problem is addressed in the following
sections. It can be structured into two parts: Few-body and many-body
problem. In our formalism, both parts find a unified description. In
Sect.\ \ref{sec:SuperfluidityPhaseTransition} we discuss the phase transition
to superfluidity at low $T$ within our formalism. We present results for the
temperature dependence of the gap and the correlation length for $T\leq
T_{\mathrm{c}}$. For a comparison with experiment, we need to relate the
microscopic parameters appearing in the microscopic action to physical
observables as the scattering length or the molecular binding energy. The
latter quantities concern the behavior of a two-atom system. In our formalism,
this can be described as the vacuum limit where density and temperature
approach zero. This is discussed in
Sect.\ \ref{Vacuum}. Sect.\ \ref{sec:phfluct} extends the simplest truncation
(discussed in Sect.\ \ref{Method}) in order to include the particle-hole
fluctuations \cite{FSDW08}. In Sect.\ \ref{sec:ClosePt}, we demonstrate the
capacity of the model to incorporate directly the universal critical physics
at the phase transition. We discuss the behavior of the scattering length for
molecules as the critical temperature is approached. In
Sect.\ \ref{sec:Density} we address the question of a quantitatively precise
computation of the density. The density sets the only scale in the unitarity
limit where the scattering length for atoms diverges. It enters directly the
theoretical predictions as the ratio $T_{\mathrm{c}}/T_{\mathrm{F}}$, where
the Fermi temperature $T_{\mathrm{F}}$ is given by the density. The
determination of the density by a direct computation of the thermodynamic
potential on the chemical potential $\mu$ leads to a lower value than the
``free particle density''. As a result, the value of
$T_{\mathrm{c}}/T_{\mathrm{F}}$ is somewhat higher. We present conclusions in
Sect.\ \ref{sec:Conclusions}.

\section{Functional Renormalization}
\label{Method}

\subsection{Effective Action and Potential}
\label{sec:EffAct}
We express our microscopic model in terms of the action
\begin{eqnarray}\nonumber 
  &&\hspace{-1.2cm} 
  S[\varphi,\psi]=\int_0^\beta d\tau \int d^3 x \Bigl(
  \psi^\dagger \left(\partial_\tau 
    -\Delta- \mu \right)\psi\\\nonumber 
  &&\hspace{1.2cm}  +\varphi^*  \left(\partial_\tau 
    -\frac{1}{2}\Delta+\nu - 2 \mu \right)\varphi\\
  &&\hspace{1.2cm}- \frac{h_\varphi}{2} 
  \left(\varphi^* \psi^T\epsilon \psi-\varphi \psi^\dagger 
    \epsilon \psi^*
  \right)\Bigr)\,,
  \label{eq:action} 
\end{eqnarray}
where $\psi^T\epsilon \psi=2\psi_1\psi_2$, $\psi^\dagger\epsilon \psi^*=2\psi_1^*\psi_2^*$ and $\triangle$ is the Laplacian.
This model describes nonrelativistic fermions in two
different states, a composite bosonic field and a Yukawa type
interaction between them. We represent the fermion field by the
complex two-component spinor $\psi=(\psi_1,\psi_2)$ ($\epsilon$ is the totally antisymmetric symbol) and the bosonic
complex scalar field by $\varphi$. Depending on the situation,
$\varphi$ can describe a bound state of two fermions (molecule), the
closed channel state of a Feshbach resonance or a Cooper-pair like
collective state of two fermions. As long as its dynamics plays no
role, it can also be considered as an auxiliary field that is used as
an effective parameterization of a purely fermionic interaction.

We work with the formalism of quantum statistics in the grand
canonical ensemble with chemical potential $\mu$. In addition to the
spatial position $\vec{x}$, the fields depend on Euclidean time
$\tau$. At nonzero temperature, this Euclidean time is wrapped around
a torus of circumference $\beta=\frac{1}{T}$. This implies for the
boson field $\varphi(\vec{x},\beta)=\varphi(\vec{x},0)$, while the
fermion field gets an additional minus sign due to its Grassmann
property $\psi(\vec{x},\beta)=-\psi(\vec{x},0)$. Here, and in the
following we choose our units such that $\hbar=k_{\mathrm{B}}=1$. In addition,
we also rescale time, and the fields in units of the fermion mass $M$
such that effectively $2M=1$. App. A gives a brief summary how
eV units for the physical observables are recovered from our units.

At tree level, the Yukawa term leads to an effective four fermion
interaction $\psi_1\,\psi_2\rightarrow\psi^\dag_1\,\psi^\dag_2$, with coupling
strength
\begin{equation}
\lambda_{\psi,\text{eff}}=-\frac{h_\varphi^2}{\nu-2\mu+iq_0
+\frac{1}{2}\vec{q}^2}.
\label{eq:lambdapsitree}
\end{equation}
Here, $q_0$ and $\vec{q}$ denote the center of mass Euclidean
frequency and momentum of the scattering particles $\psi_1$ and
$\psi_2$. The real time frequency $\omega$ obtains by analytic
continuation, $iq_0\to -\omega$, such that the microscopic dispersion
relation for the molecules is $\omega=\vec q^2/2+\nu-2\mu$. In the
limit $h^2_\varphi\to\infty,\nu\to\infty,h^2_\varphi/\nu\to~\text{const.}$
one can neglect $q_0,\vec q^2$ and $\mu$ in Eq. \eqref{eq:action},
such that the effective fermionic interaction becomes
pointlike, cf. \cite{Diehl:2005ae} and Sect. \ref{UVUniv}.

Our aim is the computation of the quantum effective action
$\Gamma[\varphi,\psi]$. It has a structure similar to the microscopic
action $S$, but with microscopic couplings and fields replaced by
renormalized macroscopic couplings and fields. Furthermore, the
couplings will depend on momentum, and $\Gamma$ contains additional
couplings not present in $S$. The quantum effective action generates
the one-particle-irreducible vertices (amputated scattering vertices)
and (inverse) full propagators. The field equations derived from an
extremum principle for $\Gamma$ are exact equations - all contributions
from quantum and thermal fluctuations are included. All scattering
information can be easily extracted from $\Gamma$ as well as all
thermodynamic relations.

For constant scalar fields $\varphi(\vec x,\tau)=\varphi$ and
$\psi(\vec x, \tau)=0$, the effective action reduces to the effective
potential $U(\varphi)$, since $\Gamma=(V_3/T)U(\varphi)$ with $V_3$
being the volume. The $U(1)$ symmetry of global phase rotations
implies that $U$ can only depend on the invariant
$\rho=\varphi^*\varphi$. The field equation $\partial
U/\partial\varphi=0$ determines the location $\varphi_0$ of the
minimum of $U$. Any nonzero $\varphi_0$ amounts to a spontaneous
breaking of the $U(1)$ symmetry. The associated Goldstone boson is
responsible for superfluidity. Therefore, $\varphi_0$ is an order
parameter -- more precisely, $2 \rho_0= 2 \varphi^*_0\varphi_0$ is the
superfluid density. The system is in the superfluid phase if
$\rho_0(T,\mu)>0$, and in the disordered or symmetric phase if
$\rho_0(T,\mu)=0$. The knowledge of $U(\rho;T,\mu)$ thus permits
a straightforward extraction of the phase diagram. Furthermore, the
value of $U$ at the minimum corresponds to the pressure
\begin{equation}\label{2A}
p(T,\mu)=-U(\rho_0;T,\mu). 
\end{equation}
(We choose a free additive constant in $U$ such that $p(0,0)=0.)$ The
grand potential or Landau thermodynamic potential $\Phi_G$ is related
to $U(\rho_0)$ by
\begin{equation}
 \Phi_G=V_3 U(\rho_0),
\end{equation}
and we obtain the usual relations for the particle density $n$,
entropy density $s$ and energy density $\epsilon$:
\begin{eqnarray}\label{2B}
\nonumber
n&=&\frac{\partial p}{\partial \mu}{\bigg |}_{T}=-\frac{\partial U}{\partial \mu}(\rho_0),\\
\nonumber
s&=&\frac{\partial p}{\partial T}{\bigg |}_{\mu}=-\frac{\partial U}{\partial T}(\rho_0),\\
\epsilon&=&-p+Ts+\mu n = U(\rho_0)+Ts+\mu n.
\end{eqnarray}

\subsection{Flow equation for the average potential}
\label{sec:floweq}

The effective potential $U$ contains all the essential information for
the description of the equilibrium properties of a homogeneous
system. Its computation is, however, a challenge, since quantum and
thermal fluctuations on all momentum scales have to be computed. The
average potential $U_k$ includes only the effects of bosonic fluctuations with
momenta $\vec q^2>k^2$ and fermionic fluctuations with $|\vec q^2-p_{\mathrm{F}}^2|>k^2$, with $p_{\mathrm{F}}$ the Fermi momentum. It is therefore a type of coarse-grained free energy. If $k$ is large enough, the complicated infrared physics for bosons and physics close to the Fermi surface for fermions is not yet included, and one expects that the complexity of the problem is
reduced. In the limit $k\to\infty$, no fluctuation effects are included
at all, and the average potential becomes the classical potential
extracted from \eqref{eq:action},
\begin{equation}\label{2c}
U_{k\to\infty}(\rho)=U_\text{cl}(\rho)=(\nu-2\mu)\rho, \quad \rho=\varphi^\ast
\varphi.  
\end{equation}
On the other hand, for $k\to 0$ all fluctuation effects are included
and $U_{k=0}$ equals the effective potential $U$. The challenge
therefore is to construct an interpolation from the microscopic
potential $U_{k\to\infty}$ to the effective potential $U_{k\to 0}$.

The dependence of $U_k$ on the ``average scale'' $k$ obeys an exact
renormalization group equation or flow equation \cite{Diehl:2007th}, with $q = (\omega , \vec q)$
\begin{eqnarray}\label{2D}
\partial_k U_k(\varphi)&=&\frac12 \int_{ q}%
\Big\{ \text{tr}_\varphi
[\bar G^{-1}_\varphi (\varphi;q)+R^{(\varphi)}_k(q)]^{-1}\partial_k R^{(\varphi)}_k(q)\nonumber\\
&& -\text{tr}_\psi [G^{-1}_\psi(\varphi;q)+R^{(\psi)}_k(q)]^{-1}\partial_k R^{(\psi)}_k(q)\Big\}.\nonumber\\
\end{eqnarray}
\begin{figure}
\begin{minipage}{\linewidth}
\begin{center}
\setlength{\unitlength}{1mm}
\begin{picture}(82,57)
\put (0,0){
    \makebox(80,56){
\begin{picture}(80,56)
      \put(0,0){\epsfxsize80mm \epsffile{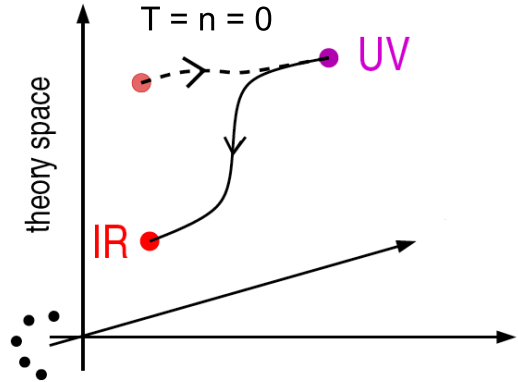}}
\end{picture}
      }}
   \end{picture}
\end{center}
\vspace*{-1.25ex} \caption{Flow of the effective action in the space of
  couplings. The flow starts at an initial scale $\Lambda$, where all
  fluctuations are suppressed and the effective action reduces to the
  classical action. By solving the FRG equations we follow the flow in
  parameter space down to $k=0$. Two-body scattering observables are measured
  in the physical vacuum defined as the state with $n=T=0$, but at $k=0$ --
  all vacuum fluctuations are included. The classical action at $k=\Lambda$ is
  chosen such as to match the two-body observables in the physical vacuum
  state (dashed line). We then switch on temperature and density to solve the
  many-body problem. The many-body flow deviates from the vacuum flow when the
  cutoff wavelength $k^{-1}$ becomes comparable to the characteristic
  thermodynamic wavelengths, i.e. the de Broglie wavelength
  $\lambda_{\text{dB}} \sim T^{-1/2}$ and the mean interparticle spacing
  $d\sim n^{-1/3}$ (solid line). }
\label{EffActFlow}
\end{minipage}
\end{figure}

The first piece accounts for the bosonic fluctuation and the second for the
fermions. Eq.\ \eqref{2D} is an exact functional differential equation. It includes all orders in a perturbative expansion as well as all non-perturbative effects. 

In order to understand the structure of this equation we first include only the bosonic contribution. Our model reduces in this case to a gas of bosons, for which the flow equations \cite{FloerchingerWetterich} and thermodynamics \cite{FWThermod} have been discussed extensively in presence of a repulsive interaction and for arbitrary dimension. In Eq.\ \eqref{2D} we encounter the propagator $\bar G_\varphi$ of the scalar field in an arbitrary background $\varphi$. We
use a two-component basis of real fields $\varphi_1,\varphi_2$, related to the complex field by $\varphi(x)=\frac{1}{\sqrt{2}}\big(\varphi_1(x)+i\varphi_2(x)\big)$. Correspondingly, $\bar G^{-1}_\varphi$ is a $2\times 2$ matrix and $\text{tr}_\varphi$ denotes the
trace in this space. In the classical approximation, the inverse propagator
reads
\begin{equation}\label{2E}
\bar G^{-1}_{\varphi,\cl}=
\left(\begin{array}{ccc}
\vec q^2/2+\nu-2\mu&-q_0\\
q_0&\vec q^2/2+\nu-2\mu
\end{array}\right).
\end{equation}
In the absence of an interaction between the bosons, it is independent of $\varphi$. For $\nu=2\mu$, the classical propagator $\bar G_{\varphi,cl}$ becomes singular for $q=(q_0,\vec q)\to 0$. It is regularized
by adding a momentum-dependent infrared cutoff $R^{(\varphi)}_k(q)$
(a unit matrix in $\varphi_1,\varphi_2$ space) which has the general properties 
\begin{eqnarray}\label{2F}
\lim_{\frac{k^2}{q^2}\to\infty}R_k(q)\sim k^2&,&\lim_{\frac{k^2}{q^2}\to\ 0}R_k(q)\to 0.
\end{eqnarray}
For $k>0$ the combination $(\bar G_\varphi^{-1}+R_k^{(\varphi)})^{-1}$ remains finite even for $\vec q^2/2+\nu-2\mu=0$. In principle, Eq.\ \eqref{2D} holds for arbitrary cutoff functions $R_k(q)$ that fulfill Eq.\ \eqref{2F}. For approximate solutions of Eq.\ \eqref{2D} an appropriate choice is important, however, and we discuss our choice of $R_k$ in App. \ref{sec:Cutoff}. 

The structure of the flow equation \eqref{2D} already becomes clear in a perturbative one loop 
truncation: if we replace $\bar G_\varphi$ by
$\bar G_{\varphi,\cl}$ in the flow equation, the inverse propagator is
independent of $k$ such that the $k$ integration can be performed
analytically
\begin{equation}\label{2G}
U_k=U_{\cl}+\frac12\int_{q_0}\int_{\vec q}\text{tr}_\varphi\ln[\bar G^{-1}_\varphi+R_k]. 
\end{equation}
With $\int_{\vec q}=\int d^3\vec q/(2\pi)^3,~\int_{q_0}=T\sum_n,~q_0=2\pi n
T$, we recognize the regularized one-loop contribution. For $k=0$ and
$\nu=2\mu$, one rediscovers the pressure of a gas of free bosons. For $T<T_{\text{c}}$,
this describes Bose-Einstein condensation.

The exact flow equation is obtained if we perform a ``renormalization group improvement'' of the propagator and replace in \Eq{2D} the classical inverse propagator by the full inverse propagator in the presence of a ``background field'' $\varphi$. The full inverse propagator is given by the second functional derivative of the average action $\Gamma_k$ with respect to the background fields $\varphi(x)$. This average action generalizes the effective potential in the presence of derivative terms for the fields. We will discuss this in more detail below. What is important at this stage is that $\bar G_\varphi^{-1}$ typically depends on $\varphi$, thus generating a non-trivial $\varphi$-dependence in the flow equation \eqref{2D}. Indeed, in the presence of effective bosonic interactions $\Gamma_k$ will not be quadratic in $\varphi$ anymore. 

If one neglected the $k$-dependence of $\bar G_\varphi^{-1}$, Eq.\ \eqref{2G}
would give the one-loop perturbative result for interacting bosons. However,
beyond this approximation $\bar G_\varphi^{-1}$ will depend on $k$ since only
fluctuations with momenta $\vec q^2>k^2$ are included in its computation. Then
Eq.\ \eqref{2D} becomes a non-linear differential equation that has to be
solved numerically. Nevertheless, its structure always remains a one-loop
equation with only one momentum integration. The only difference from one loop
perturbation theory arises from the use of the full field and momentum
dependent propagator. Higher (perturbative) loops as well as non-perturbative
effects are created due to the non-linear nature of the flow equation.

In the BEC-limit the fermion chemical potential $\mu$ assumes negative values, as appropriate for a gas of bosons. Starting from the action in Eq.\ \eqref{eq:action} we could perform the Gaussian functional integral for the fermions, resulting in an effective action involving only the bosonic field $\varphi$. In general, the effective bosonic action is a complicated non-local object. However, for large enough negative $\mu$ (as compared to the typical frequencies $\omega$ and kinetic energies $\vec q^2/2$) the bosonic action becomes effectively local. We could then use the purely bosonic flow equation, starting with a non-polynomial classical potential $U_\Lambda(\varphi)$. We therefore expect to recover in the BEC-limit many features of a boson gas with repulsive interactions. (There may be quantitative differences as compared to the case of a pointlike interaction between two bosons, due to the non-polynomial $U(\varphi)$ which also depends on $\mu$.)

Away from the BEC-limit the fermion fluctuations will play a crucial role -- they dominate in the BCS-limit. We therefore have to solve the full Eq.\ \eqref{2D}. The fermionic contribution in Eq.\ \eq{2D} is of
a similar nature as the bosonic contribution, with the characteristic minus sign for a fermion
loop and $G^{-1}_\psi$ being the full fermionic inverse
propagator. The fermionic cutoff $R^{(\psi)}_k$ is chosen such that it
regularizes the momenta close to the Fermi surface rather than the
small momenta, see App. \ref{sec:Cutoff}.

The BCS limit obtains for $\mu>0$ and large positive $\nu$ and $h_\varphi^2/\nu$. The qualitative features can now be understood in the limit where the boson fluctuations can be neglected. The inverse fermion propagator $G_\psi^{-1}$ will contain a piece $\sim h_\varphi \varphi$ due to the Yukawa interaction in Eq.\ \eqref{eq:action}. This induces contributions to the effective potential $U(\varphi)$, which lead to a minimum away from $\varphi=0$ if the temperature is low enough. In the classical approximation the fermion inverse propagator reads
\begin{equation}
G^{-1}_\psi =\left(\begin{array}{ccc}
-h_\varphi \epsilon \varphi^\ast && iq_0-(\vec q\,^2 -\mu)\\
iq_0+\vec q\,^2 - \mu && h_\varphi \epsilon \varphi
\end{array}\right).
\label{eq:1XY}
\end{equation}
This is a $4\times 4$ matrix, where we have suppressed the spinor indices. (The symbol $\epsilon_{\alpha\beta}$ acts on the spinor indices $\alpha=1,2$.)  In the limit where the $k$-dependence of $h_\varphi$ is neglected we can again integrate in order to obtain the one loop result for $k=0$
\begin{eqnarray}\label{eq:unull}
U_0(\varphi) &=& U_\text{cl}(\varphi) - \frac{1}{2} \int_{q_0} \int_{\vec q} \text{tr}_\psi \ln G_\psi^{-1}\\
\nonumber
&=& (\nu_\Lambda-2\mu) \varphi^*\varphi \\
\nonumber
&-& 2 T \int_{\vec q} \ln\left[\text{cosh}\left(\sqrt{(\vec q^2-\mu)^2+h_\varphi^2 \varphi^*\varphi}/2T\right)\right].
\end{eqnarray}
Condensation occurs if the minimum of $U_0$ occurs for $|\varphi_0|>0$. This always happens for small enough temperature $T$. Determining from Eq.\ \eqref{eq:1XY} the critical temperature $T_{\mathrm{c}}$ for the onset of a condensate $\varphi_0\neq0$ yields the BCS value. (For this purpose we relate $\nu_\Lambda$ to the scattering length $a$ in Sect.\ \ref{Vacuum}.) For quantitative accuracy the bosonic fluctuations cannot be neglected even deep in the BCS region. We will see in Sect.\ \ref{sec:phfluct} that they induce in our formalism the corrections from the particle-hole fluctuations.  

From Eq. \eqref{eq:unull} we can read off the fermion dispersion $E_{\vec q}=\sqrt{(\vec q^2-\mu)^2+h_\varphi^2 \varphi^*_0\varphi_0}$. Thus, $\Delta = h_\varphi \varphi_0$ represents the $U(1)$ symmetry breaking mass term or gap for the single particle excitations. In our framework, it is proportional to the bosonic field expectation value, such that spontaneous symmetry breaking $\rho_0 = \varphi_0^*\varphi_0\neq 0$ and a gap for the fermions coincide. The actual values of $h_\varphi, \rho_0$ are, however, determined beyond the mean field approximation, cf. e.g. Eq. \eqref{2M}.

\subsection{Truncation}
\label{sec:trunc}

The full propagators $G_{\varphi,\psi}$ remain complicated objects. It is at
this level that we need to proceed with an approximation by making an ansatz
for their general form. This amounts to a truncation of the most general form
of the average action $\Gamma_k$, which is the generalization of the average
potential $U_k$ to a functional of space- and time-dependent fields
$\varphi(x),\psi(x)$. In other words, $\Gamma_k$ is the quantity corresponding
to the effective action $\Gamma$ in the presence of the infrared cutoff $R_k$,
which suppresses the fluctuation effects arising from momenta $\vec
q^2<k^2$. 

A truncation of $\Gamma_k$ should respect the symmetries, which we discuss in App.
\ref{sec:symmetries}. Since fluctuations for the derivative terms will turn out to be important for the composite bosonic field we have to distinguish carefully between
  microscopic (bare) and renormalized fields and couplings. We use in this
paper the lowest order in a derivative expansion which, in terms of bare quantities, reads
\begin{eqnarray}\label{2H}
  \Gamma_k&=&\int_{\tau,\vec x}
  \Big\{\bar U_k(\bar \rho)+\bar \varphi^\ast(Z_\varphi\partial_\tau-\frac{1}{2}A_\varphi
  \Delta)\bar\varphi\\
  &&+\psi^\dagger(\partial_\tau-\Delta-\mu)\psi
  -\frac{\bar h_\varphi}{2}(\bar\varphi^\ast\psi^T\epsilon\psi-\bar\varphi\psi^\dagger
  \epsilon\psi^\ast)\Big\}.\nonumber 
\end{eqnarray}
The bar denotes bare fields and couplings, which can be related
  to renormalized quantities by absorbing the running gradient
  coefficient $A_\varphi$ by a field rescaling
  $\varphi=A_\varphi^{1/2}\bar \varphi$ (wavefunction
  renormalization). As we neglect the fermionic wave function
  renormalization in this work, there is no need to distinguish
  between bare and renormalized fermion fields here. In terms of
  renormalized fields, the truncation then reads
\begin{eqnarray}
\label{eq:renormalizedtruncation}
  \Gamma_k&=&\int_{\tau,\vec x}
  \Big\{U_k(\rho)+\varphi^\ast(S_\varphi\partial_\tau-\frac{1}{2}\Delta)\varphi\\
  &&+\psi^\dagger(\partial_\tau-\Delta-\mu)\psi
  -\frac{h_\varphi}{2}(\varphi^\ast\psi^T\epsilon\psi-\varphi\psi^\dagger\epsilon\psi^\ast)\Big\}.\nonumber 
\end{eqnarray}
with the renormalized couplings $S_\varphi=Z_\varphi/A_\varphi$,
$h_\varphi=\bar h_\varphi/\sqrt{A_\varphi}$, order parameter $\rho=A_\varphi \bar \rho$, and effective potential $U(\rho)=\bar U(\bar \rho)$.
In addition to the average potential $\bar U_k$, the wave function renormalizations
$Z_\varphi,A_\varphi$ and the Yukawa coupling $\bar h_\varphi$ depend on $k$. In
this approximation the renormalized inverse boson propagator $G_\varphi^{-1}=\bar G_\varphi^{-1}/A_\varphi$ reads
\begin{eqnarray}
G^{-1}_\varphi&=&\left(\begin{array}{ccc}
U'_k+2\rho U''_k+\vec q\,^2/2\,&&-S_\varphi q_0\\
S_\varphi q_0&&U'_k+\vec q^2/2
\end{array}\right).\label{2I}
\end{eqnarray}
Here, primes denote derivatives with respect to $\rho$. The fermionic
inverse propagator is given in Eq.\ \eqref{eq:1XY}, where we use a basis $(\psi,\psi^\ast)$. 

We emphasize that $U_k$ appears on the right-hand side of \Eq{2D}: at any given scale $k$ the
flow is determined by the average potential $U_k$ at the same scale $k$ rather
than by the classical potential. The same holds for the $k$-dependent Yukawa
coupling $h_\varphi$. The right-hand side of the flow equation is only
sensitive to the physics at the scale $k$. In particular, the structure of the flow equations implies that the momentum integrals in \Eq{2D} are dominated by a small range $\vec q^2\approx k^2$. This is the main reason why a derivative expansion often works rather well. In principle, $A_\varphi,Z_\varphi$ and $h_\varphi$ are momentum-dependent functions. If
this momentum dependence is not too strong, and the effective range of momenta
in the integral \eqref{2D} remains rather small, a momentum-independent
approximation is a good starting point. The truncation Eq. (\ref{eq:renormalizedtruncation}) forms the basis for the results presented in this paper. A similar truncation has been advocated for attractive fermions on the lattice in Ref.\ \cite{Strack08}. In Sect. \ref{sec:phfluct} we extend it to assess the effect of particle-hole fluctuations.

The truncation in Eq.\ \eqref{2H} has been extended in ref.\ \cite{FSW10} where renormalization effects on the fermionic self-energy and in particular a wavefunction renormalization factor for the fermions have been included. For a detailed discussion we refer to ref. \cite{FSW10}.

\subsection{Non-perturbative flow equation}
\begin{figure}
\begin{minipage}{\linewidth}
\begin{center}
\setlength{\unitlength}{1mm}
\begin{picture}(82,47)
\put (0,0){
    \makebox(80,46){
\begin{picture}(80,46) 
      \put(0,0){\epsfxsize79mm \epsffile{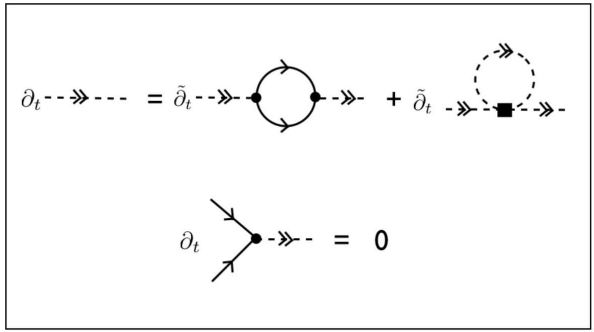}}
\end{picture}
      }}
   \end{picture}
\end{center}
\vspace*{-1.25ex} \caption{Diagrammatic representation of the flow equations
  for bare quantities in the symmetric phase. The flow of the inverse
  propagator is displayed in the first line. The frequency and momentum
  dependence of the loops is projected according to
  Eqs. (\ref{eq:projections}). The fermionic graph, present in the mean-field
  treatment taking Gaussian bosonic fluctuations into account, is supplemented
  with a feedback of bosonic fluctuations due to the second graph. These
  effects are important to capture the expected long-wavelength behavior close
  to the second-order phase transition, whereas this graph vanishes in the
  physical vacuum $n=T=0$. The Yukawa vertex in the second line does not flow
  within our truncation due the $U(1)$ symmetry. This changes in extended
  truncations \cite{FSDW08}. Moreover, in the symmetry-broken phase the flow
  equations become more complex, due to processes describing scattering off
  the condensate. In this case, also the Yukawa vertex acquires a nonzero
  flow. The $\tilde \partial_t$ derivative acts on the cutoff functions in the
  regularized propagators only. }
\label{PropFlow}
\end{minipage}
\end{figure}

For our choice of $R_k$ and with the approximation \eqref{2I}, we can perform
the momentum integration and the Matsubara sums explicitly 
\begin{eqnarray}\label{2J}
\nonumber
k\partial_kU_k &=& \eta_{A_\varphi}\,\rho\, U^\prime_k+\frac{\sqrt{2}k^5}{3\pi^2 S_\varphi} \left(1-2\eta_{A_\varphi}/5\right) s_{\mathrm{B}}^{(0)}\\
&& -\frac{k^5}{3\pi^2} \,l(\tilde \mu) \,s_{\mathrm{F}}^{(0)},\\
\nonumber
l(\tilde \mu) &=& \left(\theta(\tilde\mu+1)(\tilde \mu+1)^{3/2}-\theta(\tilde \mu-1)(\tilde \mu-1)^{3/2}\right).
\end{eqnarray}
Here we use the dimensionless chemical potential $\tilde \mu=\mu/k^2$,
the ratio $S_\varphi=Z_\varphi/A_\varphi$ and the anomalous dimension
$\eta_{A_\varphi}=-\partial \text{ln} A_\varphi /\partial \text{ln}
k$.  The \emph{threshold functions} $s_{\mathrm{B}}$ and $s_{\mathrm{F}}$ depend on
$w_1=U_k'/k^2$, $w_2=(U_k'+2\rho U_k'')/k^2$,
$w_3=h^2_\varphi\rho/ k^4$, as well as on the dimensionless
temperature $\tilde T=T/k^2$. They describe the decoupling of modes if
the effective ``masses'' or ``gaps'' $w_j$ get large. They are normalized to unity for
vanishing arguments and $\tilde T\to 0$ and read
\begin{eqnarray}\label{12K}
\nonumber
s_{B}^{(0)}&=&\left[\sqrt{\frac{1+w_1}{1+w_2}}+\sqrt{\frac{1+w_2}{1+w_1}}\right]\\
\nonumber
&&\times \left[\frac{1}{2}+N_{\mathrm{B}}(\sqrt{1+w_1}\sqrt{1+w_2}/S_\varphi)\right],\\
s_{\text{F}}^{(0)}&=&\frac{2}{\sqrt{1+w_3}}\left[\frac{1}{2}-N_{\mathrm{F}}(\sqrt{1+w_3})\right].
\end{eqnarray}
(For $s_{\mathrm{B}}^{(0)}$, only all its $\rho$ derivatives vanish for $w_1\sim w_2\to
\infty$.  
The remaining constant part is a shortcoming of the particular
choice of the cutoff acting only on spacelike momenta.)

In Eq. \eqref{12K} the temperature dependence arises through the Bose and Fermi functions
\begin{equation}
N_{B/F}(\epsilon)=\frac{1}{e^{\epsilon/\tilde T}\mp 1}.
\end{equation}
For $\tilde T \to 0$ the ```thermal parts'' $\sim N_{\mathrm{B,F}}$ vanish, whereas for
large $\tilde T$ one has 
\begin{equation}
s_{\mathrm{F}}^{(0)}\to \tilde T^{-1}, \quad s_{\mathrm{B}}^{(0)}\to 2 \tilde T S_\varphi (1+w_1)^{-1}(1+w_2)^{-1}.
\label{eq:Floweqhightemperature}
\end{equation}
In this high-temperature limit the fermionic fluctuations become unimportant. For the boson fluctuations only the $n=0$ Matsubara frequency contributes substantially. Inserting Eq. \eqref{eq:Floweqhightemperature} into Eq. \eqref{2J} yields the well known flow equations for the classical three-dimensional scalar theory with U(1) symmetry \cite{Wetterich:1992yh, Berges:2000ew}.
In App. \ref{sec:EffetivePotential} we derive the flow equations \eqref{2J} and discuss the threshold functions $s_{\mathrm{B}}$ and $s_{\mathrm{F}}$ more explicitly.

We will further truncate the general form of the average potential
\begin{eqnarray}\label{2L}
U_k&=&-p_k+m^2_{\varphi}(\rho-\rho_0)+\frac12\lambda_{\varphi}(\rho-\rho_0)^2\nonumber\\
&&-n_k \delta \mu+\alpha_k(\rho-\rho_0)\delta \mu.
\end{eqnarray}
Here $m^2_{\varphi}\equiv
m^2_{\varphi}(k),~\lambda_{\varphi}\equiv\lambda_\varphi(k),~\rho_0\equiv\rho_0(k)$ as well as
$p_k,~n_k,~\alpha_k$ depend on $k$. All couplings depend also on $T$ and a reference chemical potential
$\mu_0$ via $\mu=\mu_0+\delta \mu$. (The differentiation with respect to
$\mu$ does not act on $\mu_0$ but rather on $\delta \mu$, which we set to
zero after the $\mu$ differentiations.) In the symmetric regime, we have $\rho_0=0$, whereas the location of the minimum $\rho_0(k)$ becomes one of the parameters in the superfluid regime ($\rho_0(k)>0$). In the latter case, we have to take $m_\varphi^2=0$ such that $\rho_0(k)$ indeed corresponds to the minimum of $U_k$ for given $\mu$. We recall that $p_{k\to 0}$ is the pressure. 

An expansion of the effective potential around the minimum as in Eq.\ \eqref{2L} is expected to work reasonable for the description of second order phase transitions. In more general situations where also first order phase transitions play a role, one might use a truncation where the full function $U_k(\rho)$ is kept. The partial differential equation \eqref{2D} can then be solved numerically, for example using a discretization technique. In the present work we concentrate on the description of the second order phase transition of spin-balanced Fermi gases so that the expansion in Eq.\ \eqref{2L} is sufficient.

In the truncation of the effective potential in Eq.\ \eqref{2L}, one might include as a next step a term $\sim (\rho-\rho_0)^3$. We have not checked the influence of such a term for the BCS-BEC crossover. For a Bose gas investigated with a similar truncation \cite{FloerchingerWetterich}, the influence of such a term is quite modest, however.

The flow equations
for $p_k,m^2_{\varphi}$ or $\rho_0(k)$, and $\lambda_\varphi$ are given by
\begin{eqnarray}\label{2M}
\partial_k p_k &=& -\partial_kU_k{\big |}_{\rho_0},  \nonumber\\ 
\partial_k m^2_{\varphi} &=& \partial_kU'_k{\big |}_{\rho=0}
\quad\quad\quad\quad\quad\,\,\,\, \text{for} \quad \rho_0=0,  \nonumber\\
\partial_k \rho_0 &=& -\big(U''_k{\big
  |}_{\rho_0}\big)^{-1}\partial_kU'_k{\big |}_{\rho_0}\quad \text{for} \quad
\rho_0>0, \nonumber\\
\partial_k \lambda_{\varphi} &=& \partial_kU''_k{\big |}_{\rho_0}. 
\end{eqnarray}
Taking a derivative of Eq. \eqref{2J} with respect to $\rho$ one obtains for $\tilde T=0$
\begin{eqnarray}
\nonumber
k\partial_k U_k^\prime \!&=&\! \eta_{A_\varphi}(U_k^\prime-\rho U_k^{\prime\prime})+\frac{\sqrt{2}k}{3\pi^2 S_\varphi}\left(1-\frac{2}{d+2}\eta_{A_\varphi}\right)\\
\nonumber
&&\!\!\times\! \left[ 2\rho (U_k^{\prime\prime})^2\left(s_{\text{B,Q}}^{(1,0)}
    +3 s_{\text{B,Q}}^{(0,1)}\right)+4\rho^2 U_k^{\prime\prime} U_k^{(3)}s_{\text{B,Q}}^{(0,1)}\right]\\
&&+\frac{k}{3\pi^2}h_\varphi^2\,l(\tilde \mu)\, s_{\text{F,Q}}^{(1)}.
\label{eq:Flowu1}
\end{eqnarray}
The threshold functions $s_{\text{B,Q}}^{(0,1)}$, $s_{\text{B,Q}}^{(1,0)}$, and
$s_{\text{F,Q}}^{(1)}$ are defined in App. \ref{sec:EffetivePotential} and
describe again the decoupling of the heavy modes. They can be obtained
from $\rho$ derivatives of $s^{(0)}_{\text{B}}$ and $s^{(0)}_{\text{F}}$. Setting $\rho=0$
and $\tilde T\to0$, we can immediately infer from Eq. \eqref{eq:Flowu1} the
running of $m_\varphi^2$ in the symmetric regime.
\begin{equation}
k\partial_k m_\varphi^2=k\partial_k U_k^\prime =\eta_{A_\varphi}m_\varphi^2 
+\frac{k}{3\pi^2} h_\varphi^2\,l(\tilde \mu) \,s^{(1)}_{\text{F,Q}}(w_3=0).
\label{eq:flowmphi}
\end{equation}
One can see from Eq. \eqref{eq:flowmphi} that fermionic fluctuations lead to a
strong renormalization of the bosonic ``mass term'' $m_\varphi^2$. In the course
of the renormalization group flow from large scale parameters $k$
(ultraviolet) to small $k$ (infrared) the parameter $m_\varphi^2$
decreases strongly. When it becomes zero at some scale $k>0$ the flow
enters the regime where the minimum of the effective potential $U_k$ is at
some nonzero value $\rho_0$. This is directly related to spontaneous
breaking of the $U(1)$ symmetry and to local order. If $\rho_0\neq0$ persists
for $k\to0$ this indicates superfluidity.

For given $A_\varphi,Z_\varphi,h_\varphi$, \Eq{2J} is a nonlinear
differential equation for $U_k$, which depends on two variables $k$
and $\rho$. It has to be supplemented by flow equations for
$A_\varphi,Z_\varphi,h_\varphi$. The flow equations for the wave
function renormalization $Z_\varphi$ and the gradient coefficient
$A_\varphi$ cannot be extracted from the effective potential, but are
obtained from the following projection prescriptions,
\begin{eqnarray}\label{eq:projections}
\nonumber
\partial_t Z_\varphi &=& -\partial_t 
\frac{\partial}{\partial q_0} (\bar P_\varphi)_{12}(q_0,0){\big |}_{q_0=0} ,\\
\partial_t A_\varphi &=&2  \partial_t  \frac{\partial}{
  \partial \vec q\,^2} (\bar P_\varphi)_{22}(0,\vec q){\big |}_{\vec q=0},
\end{eqnarray}  
where the momentum dependent part of the propagator,
$\bar P_\varphi(q)=\bar G_\varphi^{-1}(q) - \bar G_\varphi^{-1}(0)$ is defined by
\begin{equation}
\frac{\delta^2 \Gamma_k}{\delta\bar \varphi_a(q)
\delta\bar \varphi_b(q^\prime)}\Big|_{\varphi_1 =
\sqrt{2\rho_0}, \varphi_2=0} = (\bar P_\varphi)_{ab}(q)\delta(q+q^\prime).  
\end{equation}
The computation of the flow of the gradient coefficient is rather involved,
since the loop depends on terms of different type, $\sim (\vec q \cdot\vec
p)^2,~ \vec q\,^2$, where $\vec p$ is the loop momentum.  The computation is
outlined in App. \ref{app:GradientCoeff}, together with the calculation of $Z_\varphi$ and the respective flow equations. In App. \ref{sec:FlowYukawaCoupling} we
describe the flow equation for the Yukawa coupling $h_\varphi$.

\begin{figure}
\begin{minipage}{\linewidth}
\begin{center}
\setlength{\unitlength}{1mm}
\begin{picture}(82,35) 
\put (0,0){
    \makebox(80,35){
\begin{picture}(80,35) 
      \put(0,0){\epsfxsize80mm \epsffile{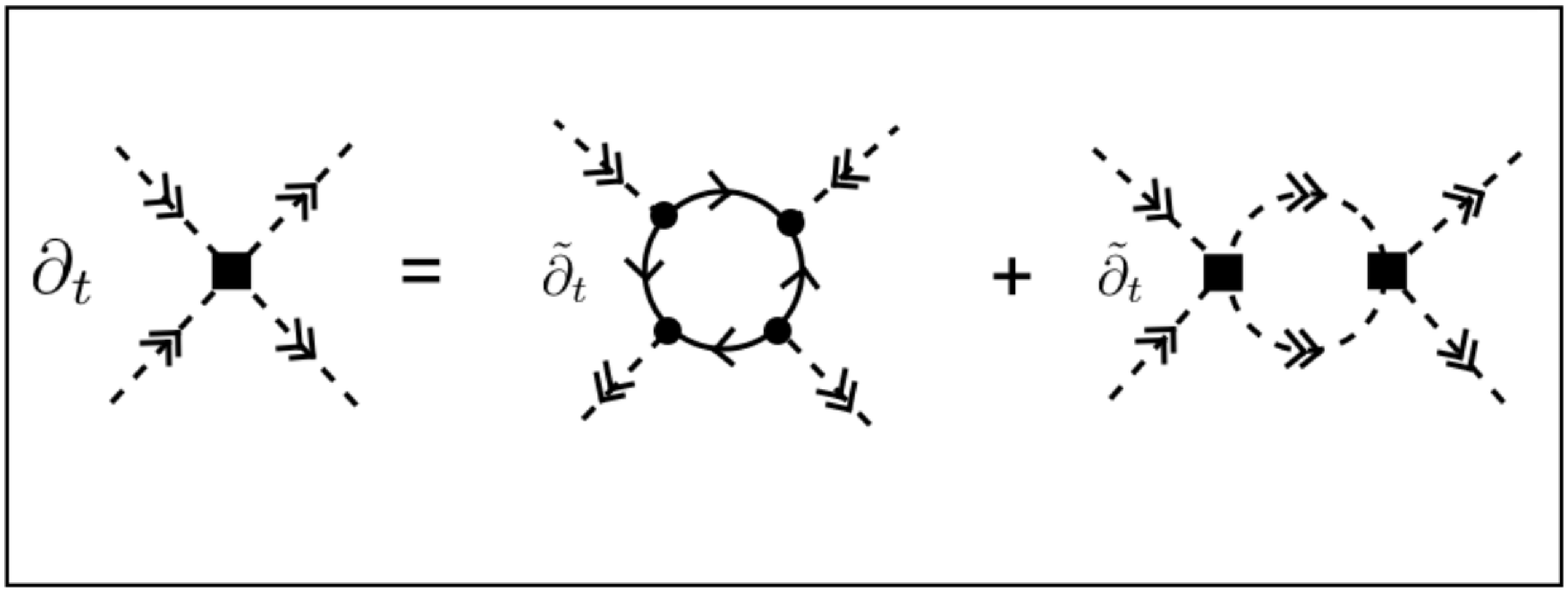}}
\end{picture}
      }}
   \end{picture}
\end{center}
\vspace*{-1.25ex} \caption{Diagrammatic representation of the flow equation
  for the bosonic self-interaction in the symmetric phase. As in
  Fig.~\ref{PropFlow}, we observe a feedback due to bosonic fluctuations. They
  have an important impact on the ratio of bosonic to fermionic scattering
  length as described in Sect. IV G extracted in the physical vacuum, as well
  as on the long-wavelength physics close to the phase transition.}
\label{IntFlow}
\end{minipage}
\end{figure}

We provide the diagrammatic representation of the flow equations considered here in the symmetric phase in Figs. \ref{PropFlow},\ref{IntFlow}. 

\subsection{Scattering length and concentration}
\label{ssec:Scatteringlengthandconcentration}

Besides the thermodynamic parameters temperature $T$ and the atom
density $n$, our model contains free parameters which characterize the
microscopic action $\Gamma_\Lambda$ and constitute the initial values
for the flow. The most important one is the ``detuning'' $\nu$, which
can be related to the scattering length $a$ for atom-atom scattering. The relation between $\nu$ and $a$ only concerns the two-body problem. We will rederive the well known function $a(\nu)$ within our formalism in Sect. \ref{Vacuum}. Relations between macroscopic quantities only involve dimensionless quantities, once appropriate units are set. We will therefore use instead of the scattering length the dimensionless concentration $c=a \kF$, where we
define the Fermi momentum $\kF$ in terms of the density,
$n=\kF^3/(3\pi^2)$. (This is a formal definition that we also use for $T\neq 0$ and in the BEC regime where no Fermi surface exists.) For a small value of $|c|$ the scattering length
is small as compared to the inter-particle distance and the system is
weakly interacting. In contrast for $|c|\gg 1$ we have to deal with a
strongly interacting system. The ``unitarity limit'' corresponds to
diverging $a$ or $c^{-1}=0$. Positive $c$ corresponds to the BEC side
of the crossover where stable molecules exist in vacuum. The BCS side
of the crossover, where the condensation phenomena are related to
``Cooper pairs'' of atoms, is described by negative $c$. In summary,
two of the parameters appearing in $\Gamma_\Lambda$, namely $\nu$ and
$\mu$, will be replaced by $a$ and $\kF$, and chosen in order to
reflect a given concentration $c$.

The microscopic model contains the Yukawa or Feshbach coupling as an
additional free parameter. We will consider the limit of a strong
Yukawa coupling, which corresponds to a broad Feshbach resonance. In
this limit the flowing Yukawa coupling approaches rapidly a partial
fixed point \cite{Diehl:2007XXX}, such that its precise microsopic
value is not relevant. For a broad Feshbach resonance, many other
possible microscopic parameters play no role either. This aspect of
universality of broad Feshbach resonances, which also holds away from
the unitrarity limit, has been extensively studied in
\cite{Diehl:2007XXX}.

To end this section let us make some remarks on the ultraviolet scale
$\Lambda$. For our numerical solution of the flow equation we have to
choose some value of $\Lambda$ but as long as it is large enough, its
precise value is not important. The reason is that for infrared cutoff
scales $k$ larger than the scales given by the temperature $k^2>T$ and
the density $k^3>n$, the flow equations have the same form as in
vacuum (i.e. $T=n=0$). In the language of quantum field theory the
vacuum model is renormalizable and the ultraviolet cutoff scale can be
chosen arbitrarily large. These arguments show that observables
expressed in dimensionless ratios, for example the gap in units of the
density $\Delta/\epsilon_F$ are independent of the choice of
$\Lambda$.

\section{Superfluidity and phase transition}
\label{sec:SuperfluidityPhaseTransition}

The presence of superfluidity is intimately connected with the
spontaneous breaking of the global continuous $U(1)$ symmetry of phase
rotations. According to Goldstone's theorem, a spontaneously
broken continuous symmetry gives rise to the existence of a massless
bosonic mode. The vanishing mass scale $m_\varphi^2 \to 0$ of this
mode in turn causes a diverging spatial correlation
length $\xi \sim 1/m_\varphi \to \infty$. This long range coherence
then provides for typical signatures of superfluidity, such as
frictionless flow and the existence of vortices as topological defects.

In our formalism, we use the spontaneous breaking of the global $U(1)$
symmetry as an indicator for the onset of superfluidity. The effective action
formalism can directly be used to classify the phases of the system according
to the symmetries of the thermodynamic equilibrium state, giving access to a
universal description of the phase transition.

In a homogeneous situation, we can consider the effective potential
$U$, which corresponds to the average potential $U_k$ for $k\to0$ or,
more precisely for $k^{-1}$ equal to the macroscopic size of the
experimental probe. The effective potential can only depend on the
invariant combination $\rho = \varphi^\ast\varphi$. The equation of
motion then reduces to a purely algebraic relation reflecting the
condition that the effective potential has an extremum at the
equilibrium value $\rho_0$,
\begin{eqnarray}\label{PotFieldEq}
  \frac{\partial U}{\partial \varphi^\ast}\Big|_{\rho_0}=
  \frac{\partial U}{\partial \rho}\Big|_{\rho_0} 
  \cdot \varphi_0 = m_\varphi^2(\rho_0) \cdot \varphi_0 = 0.
\end{eqnarray}
We have defined a ``bosonic mass term'' $m_\varphi^2$ as the
$\rho$ derivative of the effective potential. If nonzero, this mass
term provides a gap in the effective boson propagator, and thus
suppresses the boson propagation with a spatial correlation length
$\xi \sim 1/m_\varphi$. This simple equation can be used to classify the
phases of the system,
\begin{eqnarray}\label{CharPhases}
\nonumber
  \mathrm{Symmetric\,\,phase:} && \rho_0 =0, \quad m_\varphi^2 > 0,
  \\\nonumber
  \mathrm{Symmetry\,\, broken\,\, phase:} &&\rho_0  > 0, 
  \quad m_\varphi^2 = 0,\\
  \mathrm{Phase\,\, transition:} && \rho_0 = 0,\quad m_\varphi^2 = 0,
\end{eqnarray}
where $\rho_0$ denotes the solution of Eq.~(\ref{PotFieldEq}). In the
symmetric phase (SYM), we deal with a normal gas: there is no condensate,
$\rho_0=0$, and the bosonic mass does not vanish. The phase with spontaneous
symmetry breaking (SSB) is instead characterized by a nonvanishing condensate
$\rho_0>0$. This requires the vanishing of the mass term $m_\varphi^2$:
Goldstone's theorem is directly implemented in the effective action
formalism. The massless Goldstone mode is responsible for superfluidity, and
the vanishing mass is associated with a diverging spatial correlation length for the Goldstone
mode. Without loss of generality we choose the phase of spontaneous symmetry
breaking by fixing $\varphi_{1,0}=\sqrt{2\rho_0}$, $\varphi_{2,0}=0$ such that
the Goldstone mode corresponds to $\varphi_2(x)$. The second ``radial mode''
$\varphi_1(x)-\varphi_{1,0}$ describes a field with a finite correlation
length 
\begin{eqnarray}\label{corrle}
\xi_R=\sqrt{1/(2\rho_0 U^{\prime\prime}(\rho_0))}
\end{eqnarray}
(here, we have
neglected possible differences in the wave function renormalization for the Goldstone
and radial modes in the SSB regime).

The flow equation in the SYM regime
\begin{equation}\label{SYMflow}
  \partial_t m_\varphi^2 = \partial _t \frac{\partial 
    U_k}{\partial \rho}\Big|_{\rho_0=0},\quad \partial_t \rho_0 = 0,
\end{equation}
has already been discussed before, cf. Eq. \eqref{eq:flowmphi}. 
In the SSB phase, the Goldstone mass $m_\varphi^2 = (\partial U/\partial \rho)
(\rho_0) = 0$ vanishes identically. The flow equation for the
condensate follows by imposing this minimum condition at any scale in the SSB
regime,
\begin{eqnarray}\label{SSBflow}
0 &= &\frac{d}{d t} \left(\frac{\partial U_k}{
\partial \rho}\Big|_{\rho_0}\right) = \partial_t \frac{
\partial U_k}{\partial \rho}\Big|_{\rho_0} + 
\partial_t \rho_0 \frac{\partial^2 U_k}{\partial \rho^2},\nonumber\\
&&\Rightarrow\quad 
\partial_t \rho_0 = - \frac{\partial_t U'_k(\rho_0)}{\lambda_\varphi},
\end{eqnarray}
with quartic coupling
\begin{equation}
\lambda_\varphi=U_k^{\prime\prime}(\rho_0).
\end{equation}
The phase transition is characterized by the simultaneous vanishing of the
mass term and the condensate for $k\to0$. This additional constraint allows to
extract the critical temperature $T_{\text{c}}$ from Eq. \eqref{CharPhases}.

Let us discuss the emergence of spontaneous symmetry breaking in our
formalism. Independently of the strength of the interaction, at low enough
temperature the flow of the mass term (\ref{SYMflow}) hits zero at a finite
value of the flow parameter $k$. At this point we switch from
Eq. (\ref{SYMflow}) to Eq. (\ref{SSBflow}). The flow then continues in the SSB
regime, and a condensate builds up. At fixed $c = a k_\text{F}$ and for
$T<T_{\text{c}}$, the condensate saturates at some positive value $\rho_0$ for
$t\to - \infty$, i.e., when the cutoff is removed. This indicates the presence
of spontaneous symmetry breaking. In contrast, for $T>T_{\text{c}}$, the flow
drives $\rho_0$ back to zero at a finite cutoff scale $k=k_\text{SR}$. In this
case, we have to switch back to the SYM formulae \eqref{SSBflow}. The presence
of a nonzero $\rho_0(k)$ in the range $k_\text{SR}\leq k \leq k_\text{SSB}$
can be interpreted as the formation of local order on length scales between
$k_\text{SSB}^{-1}$ and $k_\text{SR}^{-1}$, which is then destroyed by
fluctuations which persist on larger length scales.

\begin{figure}
\begin{minipage}{\linewidth}
\begin{center}
\setlength{\unitlength}{1mm}
\begin{picture}(82,57)
\put (0,0){
    \makebox(80,56){
\begin{picture}(80,56)
      \put(0,0){\epsfxsize80mm \epsffile{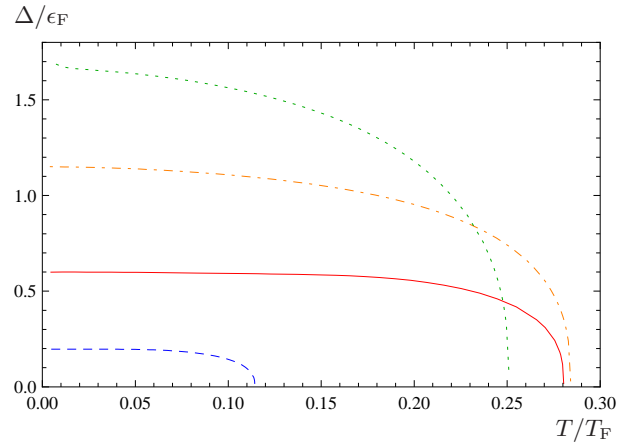}}
      \put(72,-3){$T/T_{\mathrm{F}}$}
      \put(0,52){$\Delta/\epsilon_{\mathrm{F}}$}
\end{picture}
      }}
   \end{picture}
\end{center}
\vspace*{-1.25ex} \caption{Gap in units of the Fermi energy $\Delta/\epsilon_{\mathrm{F}}=h_\varphi \sqrt{\rho_0}/\epsilon_{\mathrm{F}}$ as a function of temperature $T/T_{\mathrm{F}}$. We show the curves obtained on the BCS side with $c^{-1}=-1$ (dashed line), in the unitary regime with $c^{-1}=0$ (solid line) and on the BEC side with $c^{-1}=1$ (dashed-dotted line) and $c^{-1}=2$ (dotted line). As expected the gap for the fermions becomes larger on the BEC side for small temperatures. The results clearly establish the second order nature of the phase transition.}
\label{fig:Gap}
\end{minipage}
\end{figure}

The scale $k_\text{SSB}$ where SSB occurs in the flow varies strongly
throughout the crossover: It is instructive to discuss SSB in both the BEC and
the BCS regime. In the BEC region, SSB appears already at a high scale
$k$. The condensate builds up quickly during the flow. At very low temperature
$T \approx 0$, it represents the dominant contribution to the particle
density, apart from a tiny condensate depletion which is consistent with a
phenomenological Bogoliubov theory for bosonic molecules (described by a
scalar field $\varphi$), with a four-boson interaction extracted from the
solution of the vacuum problem. In the next section we will compute the
effective molecule scattering length $a_\text{M}$ and we find $a_{\text{M}}
=0.72 a$. This demonstrates the relevance of the inclusion of bosonic vacuum
(or quantum) fluctuations -- omitting them would result in a Bogoliubov theory
with $a_{\text{M}} = 2a$. At higher temperature, we find that \emph{bosonic
  fluctuations} tend to lower the value of the condensate at smaller $k$. We
find a second order phase transition throughout the whole BEC-BCS crossover, corresponding to the
expected second order phase transition of an O(2) model. This behavior may be inferred from Fig. \ref{fig:Gap}, where the behavior of the gap parameter is studied as a function of temperature. One clearly observes the continuous closing of the gap as the critical temperature is approached. Omitting the bosonic
fluctuations, there would be no mechanism driving the condensate down quickly
enough -- the phase transition then seems to be of first order. This artefact often occurs in other approaches to the crossover problem at finite temperature \cite{Diehl:2005ae, Haussmann:2007zz, Strinati04} and is most severe in the BEC regime. This demonstrates the importance of the inclusion of effective thermal or
statistical bosonic fluctuations.

On the BCS side, condensation only appears in the deep IR flow for $k\ll \Lambda_\text{UV}$. This goes in
line with the fact that already very low energy scales like a small temperature, destroy condensation or pairing in this
regime. Condensation is indeed a tiny effect in the BCS regime. Only an
exponentially small fraction of fermions around the Fermi surface can
contribute to pairing, as seen in Fig. \ref{fig:Gap}.

\begin{figure}
\begin{minipage}{\linewidth}
\begin{center}
\setlength{\unitlength}{1mm}
\begin{picture}(82,57)
\put (0,0){
    \makebox(80,56){
\begin{picture}(80,56)
      \put(0,0){\epsfxsize80mm \epsffile{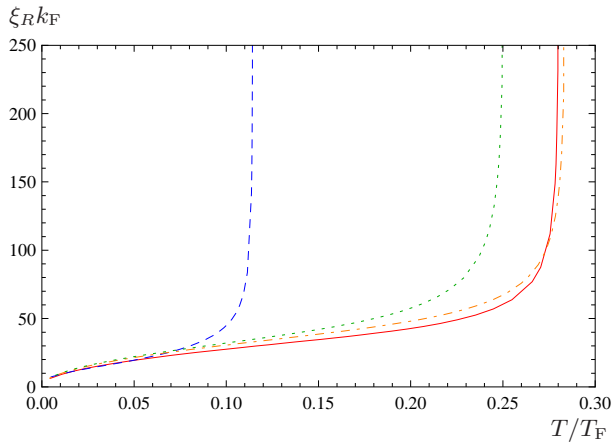}}
      \put(72,-3){$T/T_{\mathrm{F}}$}
      \put(0,52){$\xi_R \kF$}
\end{picture}
      }}
   \end{picture}
\end{center}
\vspace*{-1.25ex} \caption{Radial correlation length or healing length in units of the Fermi momentum $\xi_R \kF= (2\lambda_\varphi \rho_0)^{-1/2} \kF$ as a function of temperature $T/T_{\mathrm{F}}$. We show the curves obtained on the BCS side with $c^{-1}=-1$ (dashed line), in the unitarity regime with $c^{-1}=0$ (solid line) and on the BEC side with $c^{-1}=1$ (dashed-dotted line) and $c^{-1}=2$ (dotted line).}
\label{fig:Xi}
\end{minipage}
\end{figure}

Our construction demonstrates the emergence of an effective bosonic theory in
two respects: The first one concerns the effective bosonic theory for the
tightly bound molecules in the BEC regime. The ground state at $T=0$ is indeed
a Bose-Einstein condensate, with a small depletion due to bosonic
self-interactions. Increasing the temperature, bosonic fluctuations drive the
system back into the symmetric phase. This second step is precisely the
mechanism observed previously in purely bosonic theories
\cite{ATetradis93,BTetWet94}.  The presence of bosonic long range fluctuations is illustrated in Fig. \ref{fig:Xi}, where we plot the correlation length which relates to the bosonic mass term as indicated in Eq. \eqref{corrle}. This aspect of bosonic behavior extends over the whole crossover and is discussed in more detail in Sect. \ref{sec:ClosePt}.

In this context, we note a shortcoming of the present truncation in the bosonic sector which manifests itself in the deep infrared regime of the flow in the low temperature phase. For example, in the present truncation the four-boson coupling $\lambda_\varphi$ vanishes polynomially at zero temperature, while a more sophisticated infrared analysis for a weakly interacting Bose gas reveals a logarithmic flow of this coupling \cite{Castellani04,Wetterich08,Dupuis07}. In fact, the wave function renormalization $Z_\varphi$, i.e. the coefficient to the linear frequency term in the inverse boson propagator, flows towards zero. In a more accurate extended truncation a quadratic frequency term is generated by the renormalization group flow. The thereby modified power counting results in logarithmic rather than polynomial divergencies. The correct behavior could thus be implemented by extending the truncation with a quadratic frequency term in the inverse boson propagator \cite{Wetterich08,Dupuis07,FloerchingerWetterich}. We note, however, that our results for thermodynamic quantities are not affected by the deep infrared physics at much longer wavelengths than the de Broglie wavelength or the mean interparticle spacing. Furthermore, the critical behavior at the finite temperature phase transition cannot be affected by such an extension of the truncation, as only the Matsubara zero mode governs the behavior in this region. 
\begin{figure}
\begin{minipage}{\linewidth}
\begin{center}
\setlength{\unitlength}{1mm}
\begin{picture}(82,57)
\put (0,0){
    \makebox(80,56){
\begin{picture}(80,56)
      \put(0,0){\epsfxsize80mm \epsffile{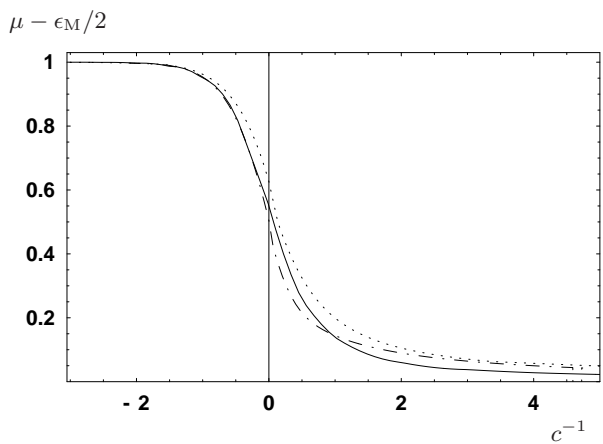}}
      \put(72,-3){$c^{-1}$}
      \put(0,52){$\mu - \epsilon_\text{M}/2$}
\end{picture}
      }}
   \end{picture}
\end{center}
\vspace*{-1.25ex} \caption{Chemical potential 
for $T=0$ minus half the binding energy
  $\epsilon_{\text{M}}/2 = -\theta(c^{-1}) c^{-2}$ as a function of the inverse concentration $c^{-1}=(ak_F)^{-1}$.
  We compare our FRG result (solid) to
  extended mean field theory (dotted; no bosonic loops) and a result based on the Schwinger-Dyson
  equations (dot-dashed) \cite{Diehl:2005ae}.}
\label{fig:ChPot}
\end{minipage}
\end{figure}

\section{Vacuum limit}
\label{Vacuum}

In this section, we consider the effective action in the vacuum limit, which
is obtained from $\Gamma_{k\to0}$ in the limit $n\to0$, $T\to0$.  In this
limit, the effective action generates directly the basic building blocks for
the amputated $n$-point scattering amplitudes, that in turn directly yield the
cross section \cite{Diehl:2005ae}. Only simple tree diagrams have to be
evaluated in order to compute, for example, the two-atom scattering via boson
exchange. For more complicated scattering observables where loop effects enter, one observes a considerable simplification of
the diagrammatic structure of the flow equations in the vacuum limit.
 
In order to make contact with experiment, we have to relate the microscopic
parameters which characterize the theory at a high momentum scale
$\Lambda_{\text{UV}}$ to the macroscopic observables for two-atom scattering
in vacuum, like the scattering length $a$ or the molecular binding
energy. This is the analogue of renormalization in quantum electrodynamics,
which relates the bare (microscopic) to the renormalized (macroscopic)
coupling. For our setting the bare couplings diverge in the limit of infinite
UV cutoff $\Lambda_\text{UV}$, while all observable quantities remain finite
once they are expressed in terms of the renormalized parameters. In our
approach, this mapping between bare and renormalized parameters is implemented
by choosing the initial parameters at the microscopic scale
$\Lambda_{\text{UV}}$ such that a few two-body observables in vacuum are
matched to their experimental values in the limit $k\to 0$.

In our picture, atom scattering in vacuum is mediated by the formation and
dissociation of a collective boson. We stress, however, that in the limit of
broad resonances, $\bar h_\varphi\to\infty$, the collective bosonic degree of
freedom is purely auxiliary on the level of the microscopic action (defined at
$\Lambda_{\text{UV}}$) \cite{Diehl:2005ae}. In this case the microscopic interaction is simply a
pointlike contact interaction. Nevertheless, fermionic fluctuations
dynamically generate a physical bosonic bound state on the BEC side of the
resonance where the fermionic scattering length is positive. For $k\to 0$, we
find bosonic degrees of freedom with dispersion $\omega = q^2/(4M)$ -- the
boson mass $2M$ is appropriate for the composite objects. On the BCS side
close to the Feshbach resonance (negative $a$) the bosonic particle remains as
a resonant state, with energy now somewhat above the open channel energy
level.

In the broad resonance limit and not too far away from the Feshbach resonance,
the physics of both the vacuum and the thermodynamics becomes very insensitive
to details of the microscopic parameters
\cite{Diehl:2005ae,Nishida:2006br,Diehl:2007XXX}. We find that further microscopic
information is suppressed at least $\sim \bar h_\varphi^{-2}$ for large
  $\bar h_\varphi$. In the RG language, this ``universality'' is conveniently
expressed in terms of a (nontrivial) broad resonance fixed point. As a
signature of this strong universality, the two-channel model with explicit
molecular degrees of freedom becomes equivalent to a single channel model, in
which the interaction is \emph{a priori} specified by a single scale, the
scattering length -- both models are characterized by the same fixed point.
We recover this feature directly from the solution of the flow equations.
 
The physics of the vacuum limit describes few-body scattering, which can also
be computed from quantum mechanics. The functional integral approach is
equivalent to quantum mechanics, and the flow equations are exact. The results
from a solution of the flow equations for $k\to 0$ should therefore coincide
with the quantum mechanical results, up to artifacts of the truncation. One may
employ the comparison to quantum mechanics for an estimate of the error due to
the truncation.

At first sight, the use of the flow equation machinery for the solution of a
quantum mechanical problem may seem unduely complicated. We recall, however,
that the flow equation result can immediately be extended to the many-body
problem at nonzero density or temperature, simply by changing the parameters
$\tilde \mu$ and $\tilde T$ in the flow equation \eqref{2J}. This is not
possible for quantum mechanics. Furthermore, many different microscopic
Hamiltonians can now be investigated with comparatively little effort -- it
suffices to change the initial values of the flow. The fixed-point behavior
offers an easy access to universality, which cannot be seen so directly in
quantum mechanics.

The vacuum problem can be structured in the following way: The two-body sector
describes the pointlike fermionic two-body interactions. It involves couplings up to
fourth order in the fermion field in a purely fermionic setting. In the language with a composite boson field $\varphi \sim
\psi\psi$ we have in addition terms quadratic in $\varphi$ or $\sim \psi\psi\varphi^*+h.c.$. The two-body sector decouples from the sectors involving a higher number of particles, described by
higher-order interactions parameters such as the dimer-dimer scattering (which
corresponds to interactions of eighth order in $\psi$ in a fermionic
language). This decoupling reflects the situation in quantum mechanics, where a two-body calculation (in vacuum) never needs input from states with more than two particles.

Not too far from unitarity the solution of the two-body problem fixes the
independent renormalized couplings completely. The couplings in the sectors
with higher particle number then are derived quantities which can be computed
as functions of the parameters of the two-body problem. Here, we will study
the dimer-dimer or molecular scattering length $a_{\rm M}$ as an important
example.

We emphasize that this almost complete ``loss of microscopic memory'' near a
Feshbach resonance is characteristic for our system with two species of
fermions. For three species with SU(3) symmetry, or for bosonic Fesh\-bach
resonances, the three-body sector will involve an additional ``renormalized
coupling'', associated to the energy of a trion bound state in the $SU(3)$ case 
\cite{BraatenHammer, FSMW}. For the present system, however, we expect
that close to a (broad) Feshbach resonance all ``macroscopic'' physical
phenomena can be expressed by only one renormalized dimensionless parameter,
namely the concentration $c=a\kF$, $\kF=(3\pi^2 n)^{1/3}$, plus the dimesionless
temperature. The inverse concentration $c^{-1}$ is a relevant parameter in the
sense of critical phenomena, with $c^{-1}=0$ being the location of the fixed
point. Scales may be set by the density $n$ or the scattering length $a$.

\subsection{Projection onto the Vacuum} 

First we specify the prescription which projects the effective action onto the
vacuum limit \cite{Diehl:2005ae}:
 \begin{eqnarray}
 \Gamma (vac) = \lim\limits_{T,\kF \to 0} \Gamma_{k = 0} \Big|_{T > T_{\text{c}} }.
\end{eqnarray}
Here $\kF$ can be viewed as the inverse mean interparticle spacing $\kF \sim
1/d$. Taking the limit $\kF \to 0$ then corresponds to a diluting procedure
where the density of the system becomes arbitrarily low. However, the limit is
constrained by keeping the dimensionless temperature $\tilde T = T/\epsilon_{\mathrm{F}}
= 2M T/\kF^2$ above criticality. This ensures that many-body effects such as
condensation phenomena are absent. In this limit, our functional integral approach becomes equivalent to quantum mechanics.

We find that for $n=T=0$ the crossover at finite density turns into a
second order ``phase transition'' in vacuum \cite{Diehl:2005ae,Nishida:2006br} as a function
of $\bar m_{\varphi }^2(\Lambda)$ which is given, in turn, by the magnetic
field $B$. This is expressed by the following constraints which separate the
two qualitatively different branches of the physical vacuum,
(cf. Eq.~\eqref{CharPhases}) 
\begin{eqnarray}\label{VacCond}
  \begin{array}{l l l}
    { \bar m_\varphi^2 |_{k\to0}>0, \quad \mu = 0  }& 
    \text{atom phase}
    & (a^{-1} < 0)  \\
    {\bar m_\varphi^2|_{k\to0} = 0,\quad \mu < 0   }& 
    \text{molecule phase}
    & (a^{-1} > 0)   \\
    {\bar m_\varphi^2|_{k\to0} = 0,\quad \mu = 0  }& 
    \text{resonance}
    & (a^{-1} = 0)
\end{array}.
\end{eqnarray}
These formulae have a simple interpretation: On the BCS side, the bosons
experience a gap $\bar m_\varphi^2>0$ and the low-density limit describes only
fermionic atoms. On the BEC side, the situation is reversed: the fermion
propagation is suppressed by a gap $-\mu$. The ground state is a
stable molecule, and the fermionic chemical potential can be interpreted as
half the binding energy of a molecule, $\epm = 2 \mu$
\cite{Diehl:2005ae} -- this is the amount of energy that must be given to a
molecule to reach the fermionic scattering threshold.

Evaluating the bosonic mass term $\bar m_\varphi^2$ in vacuum on the BEC side
(see below), one finds the well-known universal relation between binding
energy and scattering length in vacuum, $\epm = - 1/(Ma^2)$ in the
broad-resonance limit $\hpb\to \infty$. Close to the resonance,
$\epsilon_{\text{M}}\to0$, this holds for arbitrary $\bar h_\varphi$ as well as other
microscopic details and hence establishes the second order nature of the
vacuum phase transition.

\subsection{Diagrammatic simplifications}
\label{sec:simplif}

On the technical side, the procedure specified above leads to a
massive simplification of the diagrammatic structure as compared to
the finite density and temperature system. With the aid of the residue
theorem, we can prove the following statement: \emph{All diagrams
  with inner lines pointing into the same direction (thereby forming a
  closed tour of particle number) do not contribute to the flow in vacuum.} Here, the direction
of a line is given by the $q_0$ flow of the propagator. 

For a proof, we first consider the form of the renormalized inverse fermion and boson propagators in the vacuum
limit which there become diagonal and are represented by the entries
\begin{eqnarray}
P_{\mathrm{F}}(q) &=& i q_0 + \vec q\,^2 - \mu ,\nonumber\\
P_\varphi(q)  &=& i S_\varphi q_0 + \vec q\,^2/2 + m_\varphi^2 .
\end{eqnarray}
Lines pointing into the same direction represent integrals over products of
$P_{\mathrm{F}}^{-1},P_\varphi^{-1}$ with the \emph{same} sign of the frequency variable in the loop. Without loss of generality, we can choose it to be
positive.

In the presence of a nonzero cutoff, the spacelike real part of the
regularized inverse propagators $P_{\text{F}} + R_k^{\text F}, P_\varphi +
R_k^\varphi$, including the mass terms, is always positive. Hence, the poles
all lie in the upper half of the complex plane. Closing the
integration contour in the lower half-plane, no residues are picked up, implying that
these integrals vanish. A derivative with respect to the cutoff $R_k$
increments the number of inner lines $P_{\text F}^{-1},P_\varphi^{-1}$ by one (it
changes the multiplicity of the poles), but does not affect the sign of the
real part of the propagator nor of $q_0$, such that our argument remains
valid for both the regularized loops and their $k$ derivative entering the
flow equation.

Let us now discuss different types of diagrams explicitly which do \emph{not}
contribute to the vacuum flow.

\begin{itemize}

\item The mixed diagram with one inner fermion and one inner boson line,
  driving the renormalization of the fermion propagator, does not
  contribute. This implies for all $k$
\begin{eqnarray}
  P_{\mathrm{F}}(q) = i \omega + \vec q^2 - \mu.
\end{eqnarray}

\item The box diagram with two inner boson and fermion lines, in principle
  generating a four-fermion interaction even for vanishing initial
  coupling, does not appear. For $\lambda_{\psi}|_{k=\Lambda} =0$ we thus have
\begin{eqnarray}
  \lambda_\psi =0
\end{eqnarray}
at all scales. This means that partial bosonization is very efficient in this
limit -- fluctuations are completely absorbed into the bosonic sector, and
there is no ``backreaction'' on the fermion propagator.

\item Contributions to the flow from diagrams involving solely \emph{one}
  inner line are zero. This is directly related to the fact that the particle
  density vanishes by construction -- diagrammatically, the trace over the
  full propagator is represented as a closed loop coupled to an external
  current, the chemical potential (tadpole graph). This is relevant for the
  bosonic contribution to the boson mass which thus vanishes -- the
  renormalization of the boson mass in the physical vacuum is purely driven by
  the fermion loop.
\item The bubble diagrams (involving two inner boson or two inner fermion
  lines forming a closed tour) vanish.
\end{itemize}

The ladder diagrams instead, which have lines pointing in the same
direction, contribute. Due to the opposite signs of the momentum variables in
the propagators, the poles of these diagrams are located in both the upper and
the lower half plane. It is worth noting that the four-boson coupling receives
such a bosonic ladder contribution (cf. Fig. \ref{IntFlow}), which will be discussed in Sect. \ref{DimerDimer}.

\subsection{Solution of the two-body problem}
\label{UVUniv}

The two-body problem is best solved in terms of the bare couplings. Their flow
equations read
\begin{eqnarray}
\nonumber
\partial_k \bar{m}_\varphi^2 &=& \frac{\bar{h}_\varphi^2}{6\pi^2
  k^3}\theta(k^2+\mu)(k^2+\mu)^{3/2},\\ 
\nonumber
\partial_k Z_\varphi &=& -\frac{\bar{h}_\varphi^2}{6\pi^2
  k^5}\theta(k^2+\mu)(k^2+\mu)^{3/2},\\ 
\nonumber
\partial_k A_\varphi &=& -\eta_{A_\varphi} A_\varphi/k=
-\frac{\bar{h}_\varphi^2}{6\pi^2 k^5}\theta(k^2+\mu)(k^2+\mu)^{3/2},\\ 
\partial_k \bar{h}_\varphi &=&0.
\label{eq:vacuumflow}
\end{eqnarray}
The flow in the two-body sector is driven by fermionic diagrams only.  There
is no renormalization of the Feshbach coupling due to $U(1)$ symmetry. This
statement holds in the absence of an initial coupling,
  $\lambda_\psi|_{k=\Lambda}=0$, where the whole scattering amplitude of the
fermions is expressed by the exchange of a bosonic bound state. (Taking an
additional background coupling into account leads to a renormalization of
$\bar h_\varphi$ \cite{Diehl:2007XXX} which is compatible with the $U(1)$
invariance.)

Eqs. \eqref{eq:vacuumflow} are solved by direct
integration with the result
\begin{widetext}
\begin{eqnarray}
\nonumber
\bar{m}_\varphi^2(k) &=&
\bar{m}_\varphi^2(\Lambda)-\theta(\Lambda^2+\mu)\frac{\bar{h}_\varphi^2}{6\pi^2}{\bigg
  [}\sqrt{\Lambda^2+\mu}\,\left(1-\frac{\mu}{2\Lambda^2}\right)
-\frac{3}{2}\sqrt{-\mu}\,\,\text{arctan}
\left(\frac{\sqrt{\Lambda^2+\mu}}{\sqrt{-\mu}}\right){\bigg  ]} \nonumber\\
&&
+\theta(k^2+\mu)\frac{\bar{h}_\varphi^2}{6\pi^2}{\bigg
  [}\sqrt{k^2+\mu}\,\left(1-\frac{\mu}{2k^2}\right) 
-\frac{3}{2}\sqrt{-\mu}\,\,\text{arctan}
\left(\frac{\sqrt{k^2+\mu}}{\sqrt{-\mu}}\right){\bigg  ]}\label{eq:mvac},\\ 
Z_\varphi(k) &=&
Z_\varphi(\Lambda)-\theta(\Lambda^2+\mu)
\frac{\bar{h}_\varphi^2}{48\pi^2}{\bigg [}\sqrt{\Lambda^2+\mu}
\,\frac{\left(5\Lambda^2+2\mu\right)}{\Lambda^4}
-\frac{3}{\sqrt{-\mu}}\,\,\text{arctan}
\left(\frac{\sqrt{\Lambda^2+\mu}}{\sqrt{-\mu}}\right){\bigg  ]}\nonumber\\
&&
+\theta(k^2+\mu)
\frac{\bar{h}_\varphi^2}{48\pi^2}{\bigg [}\sqrt{k^2+\mu}
\,\frac{\left(5k^2+2\mu\right)}{k^4}
-\frac{3}{\sqrt{-\mu}}\,\,\text{arctan}
\left(\frac{\sqrt{k^2+\mu}}{\sqrt{-\mu}}\right){\bigg  ]}\label{eq:Zvac},\\ 
A_\varphi(k) &=& A_\varphi(\Lambda)-\theta(\Lambda^2+\mu)
\frac{\bar{h}_\varphi^2}{48\pi^2}{\bigg [}\sqrt{\Lambda^2+\mu}\,
\frac{\left(5\Lambda^2+2\mu\right)}{\Lambda^4}
-\frac{3}{\sqrt{-\mu}}\,\,\text{arctan}
\left(\frac{\sqrt{\Lambda^2+\mu}}{\sqrt{-\mu}}\right){\bigg ]}\nonumber\\
&&
+\theta(k^2+\mu)
\frac{\bar{h}_\varphi^2}{48\pi^2}{\bigg [}\sqrt{k^2+\mu}\,
\frac{\left(5k^2+2\mu\right)}{k^4}
-\frac{3}{\sqrt{-\mu}}\,\,\text{arctan}
\left(\frac{\sqrt{k^2+\mu}}{\sqrt{-\mu}}\right){\bigg ]}.
\label{eq:vacuumsolutions}
\end{eqnarray}
\end{widetext}
Here, $\Lambda$ is the initial ultraviolet scale. Let us discuss the initial value for the boson mass. It is given by
\begin{equation}
\bar{m}_\varphi^2(\Lambda)=\nu(B) -2\mu+\delta
\nu(\Lambda). \label{eq:new41} 
\end{equation}
It features a physical part, the detuning $\nu(B)= \mu_{\text{M}} (B-B_0)$, which describes the energy level of the microscopic state represented by the
field $\varphi$ with respect to the fermionic state $\psi$. At a
Feshbach resonance, this energy shift can be tuned by the magnetic field $B$, $\mu_{\text{M}}$ denotes the magnetic moment of the field $\varphi$, and
$B_0$ is the resonance position. Physical observables such as the scattering length
and the binding energy are obtained from the full effective action and are
therefore related to the coupling constants at the infrared scale
$k=0$. The quantity $\delta\nu(\Lambda)$ denotes the renormalization
  counter term that has to be adjusted such that the condition (\ref{VacCond})
  is satisfied for $k\to0$ on the resonance, see below. Equations similar to Eq. (\ref{eq:vacuumflow}) are derived in \cite{DSK08,Birse08}, where other cutoff functions are used, however.

\subsection{Renormalization}

We next show that close to a Feshbach resonance the ``microscopic'' parameters (initial conditions to the flow equations) $\bar  m_{\varphi,\Lambda}\equiv\bar m_\varphi^2(k=\Lambda)$ and $\bar
h_{\varphi,\Lambda}^2\equiv\bar h_\varphi^2(k=\Lambda)$ which enter our calculations are related to the physical observables, the detuning from the Feshbach resonance $\nu= \mu_{\text{M}}(B-B_0)$ and its width $\bar h_{\varphi}^2$. Alternatively to the set  $\{\nu, \bar h_\varphi^2\}$, we may also choose $\{a, \bar h_\varphi^2\}$, where $a$ is the scattering length. The idea behind the procedure of fixing the parameters in our RG framework is illustrated and discussed in Fig. \ref{EffActFlow}. The relations are
\begin{equation}
\bar m_{\varphi,\Lambda}^2=\mu_\text{M}(B-B_0)-2\mu+\frac{\bar h_{\varphi,\Lambda}^2}{6\pi^2}\Lambda, \quad \bar h_{\varphi,\Lambda} = \bar h_{\varphi}
\label{eq:mvarphiLambda}
\end{equation}
and the physical observables are related by 
\begin{equation}
a=-\frac{\bar h_{\varphi, \Lambda}^2}{8\pi \mu_\text{M}(B-B_0)}.
\label{eq:hvarphiLambda}
\end{equation}
In the vacuum limit specified above, and in the absence of background interactions, the Yukawa or Feshbach coupling $\bar h_{\varphi,\Lambda}$ is not renormalized, and so does not change with scale. Therefore, unlike the mass term which receives UV renormalization, the Feshbach coupling is a direct physical observable, associated to the width of the resonance (see below). In the limit of a broad Feshbach resonance only a single parameter is needed to describe the scattering physics -- only the ratio of the Feshbach coupling $\bar h_{\varphi, \Lambda}^2$ and the detuning is physical. 
Away from the Feshbach resonance the Yukawa coupling may depend on $B$, $\bar
h_\varphi^2(B)=\bar h_\varphi^2+c_1(B-B_0)+\dots$. Also the microscopic
difference of energy levels between the open and closed channel may show
corrections to the linear $B$-dependence,
$\nu(B)=\mu_\text{M}(B-B_0)+c_2(B-B_0)^2+\dots$ or $\mu_\text{M}\to
\mu_\text{M}+c_2(B-B_0)+\dots$. Using $\bar h_\varphi^2(B)$ and
$\mu_\text{M}(B)$ our formalism can easily be adapted to a more general
experimental situation away from the Feshbach resonance. The relations in
Eqs. \eqref{eq:mvarphiLambda} and \eqref{eq:hvarphiLambda} hold for all
chemical potentials $\mu$ and temperatures $T$. For a different choice of the
cutoff function the coefficient $\delta\nu(\Lambda)$ being the term
  linear in $\Lambda$ in Eq. \eqref{eq:mvarphiLambda} might be modified.

We want to connect the bare parameters $\bar m_{\varphi,\Lambda}^2$ and $\bar
h_{\varphi,\Lambda}^2$ with the magnetic field $B$ and the scattering length $a$ for
fermionic atoms as renormalized parameters. In our units, $a$ is related to
the effective interaction $\lambda_{\psi,\text{eff}}$ by
\begin{equation}
a=\frac{\lambda_{\psi,\text{eff}}}{8\pi}.
\end{equation}
In the absence of a background interaction, the fermion interaction
$\lambda_{\psi,\text{eff}}$ is determined by the molecule exchange process in
the limit of vanishing spatial momentum 
\begin{equation}
\lambda_{\psi,\text{eff}}=-\frac{\bar
  h_{\varphi,\Lambda}^2}{\bar{P}_\varphi(\omega,\vec{p}^2=0,\mu)}. 
\label{eqlambdaeffmoleculeexchange}
\end{equation}
Here, we work with Minkowski frequencies $\omega$ related to the Euclidean ones by $\mathrm i q_0 \to -\omega$.
Even though \eqref{eqlambdaeffmoleculeexchange} is a tree-level process, it is
not an approximation, since  $\bar{P}_\varphi\equiv\bar
  P_{\varphi}|_{k\to0}$ denotes the full bosonic propagator which includes
all fluctuation effects. The frequency in
Eq. \eqref{eqlambdaeffmoleculeexchange} is the sum of the frequency of the
incoming fermions which in turn is determined from the on-shell condition
\begin{equation}
\omega=2\omega_\psi=-2\mu.
\label{eqinitialvalueofmass}
\end{equation} 

On the BCS side we have
$\mu=0$ (see Eq. \eqref{VacCond}) and find with 
\begin{equation}
\bar P_\varphi(\omega=0,\vec q=0)=\bar m_\varphi^2(k=0)\equiv \bar m^2_{\varphi,0}
\end{equation}
the relation
\begin{equation}
\lambda_{\psi,\text{eff}}=-\frac{\bar h_{\varphi,\Lambda}^2}{\bar m_{\varphi,0}^2},
\end{equation}
where $\bar m_{\varphi,0}^2=\bar m_\varphi^2(k=0)$. For the bosonic mass terms at $\mu=0$, we can read off from
Eqs. \eqref{eq:vacuumsolutions} and \eqref{eq:new41} that 
\begin{equation}
  \bar{m}_{\varphi,0}^2=\bar{m}_{\varphi,\Lambda}^2
  -\frac{\bar{h}_{\varphi,\Lambda}^2}{6\pi^2} \Lambda 
  = \mu_{\text{M}}(B-B_0)+\delta \nu(\Lambda)
  -\frac{\bar{h}_{\varphi,\Lambda}^2}{6\pi^2}\Lambda.
\end{equation}
To fulfill the resonance condition in Eq. \eqref{VacCond} for $B=B_0$, $\mu=0$,
we choose
\begin{equation}
\delta \nu(\Lambda)=\frac{\bar{h}_{\varphi,\Lambda}^2}{6\pi^2}\Lambda.
\end{equation}
The shift $\delta\nu (\Lambda)$ provides for the additive UV renormalization
of $\bar{m}_\varphi^2$ as a relevant coupling.  It is exactly cancelled by the
fluctuation contributions to the flow of the mass. This yields the general
relation \eqref{eq:mvarphiLambda} (valid for all $\mu$) between the bare mass
term $\bar m_{\varphi,\Lambda}^2$ and the magnetic field. On the BCS side we
find the simple vacuum relation
\begin{equation}
\bar m_{\varphi,0}^2=\mu_\text{M}(B-B_0).
\end{equation}
Furthermore, we obtain for the fermionic
scattering length
\begin{equation}
a=-\frac{\bar{h}_{\varphi,\Lambda}^2}{8\pi \mu_{\text{M}}(B-B_0)}.
\label{eqscatterinlengthandmageticfield}
\end{equation}
This equation establishes Eq. \eqref{eq:hvarphiLambda} and shows that
$\bar{h}_{\varphi,\Lambda}^2$ determines the width of the resonance. We have thereby
fixed all parameters of our model and can express $\bar m_{\varphi,\Lambda}^2$
and $\bar h_{\varphi,\Lambda}^2$ by $B-B_0$ and $a$. The relations
\eqref{eq:mvarphiLambda} and \eqref{eq:hvarphiLambda} remain valid also at
nonzero density and temperature. They fix the ``initial values'' of the flow
($\bar h_\varphi^2\to \bar h_{\varphi,\Lambda}^2$) at the microscopic scale
$\Lambda$ in terms of experimentally accessible quantities, namely $B-B_0$ and
$a$.

On the BEC side, we encounter $\mu<0$ and thus $\omega>0$. We therefore need
the bosonic propagator for $\omega\neq 0$. Even though we have computed
directly only quantities related to $\bar P_\varphi$ at $\omega=0$ and
derivatives with respect to $\omega$ ($Z_\varphi$), we can obtain information
about the boson propagator for nonvanishing frequency by using the semilocal
$U(1)$ invariance described in App. \ref{sec:symmetries}. In momentum space,
this symmetry transformation results in a shift of energy levels
\begin{eqnarray}
\nonumber
\psi(\omega, \vec{p}) &\to& \psi(\omega-\delta,\vec{p})\\
\nonumber
\varphi(\omega,\vec{p}) &\to& \varphi(\omega-2\delta,\vec{p})\\
\mu &\to& \mu+\delta.
\end{eqnarray}
Since the effective action is invariant under this symmetry, it follows for
the bosonic propagator that
\begin{equation}
\bar{P}_\varphi(\omega,\vec{p},\mu)
=\bar{P}_\varphi(\omega-2\delta,\vec{p},\mu+\delta).
\end{equation}
To obtain the propagator needed in Eq. \eqref{eqlambdaeffmoleculeexchange}, we
can use $\delta=-\mu$ and find as in Eq. \eqref{eqscatterinlengthandmageticfield}
\begin{equation}
\lambda_{\psi,\text{eff}}
=-\frac{\bar{h}_{\varphi,\Lambda}^2}{\bar{P}_\varphi(\omega=0,\vec{p}^2=0,\mu=0)}
=-\frac{\bar h_{\varphi,\Lambda}^2}{\mu_\text{M}(B-B_0)}.
\end{equation}
Thus the relations \eqref{eq:mvarphiLambda} and \eqref{eq:hvarphiLambda} for
the initial values $\bar m_{\varphi,\Lambda}$ and $\bar h_{\varphi,\Lambda}^2$
in terms of $B-B_0$ and $a$ hold for both the BEC and the BCS side of the
crossover.


\subsection{Binding energy}

We next establish the relation between the molecular binding energy
$\epsilon_\text{M}$, the scattering length $a$, and the Yukawa coupling $\bar
h_{\varphi,\Lambda}^2$. From Eq. \eqref{eq:vacuumsolutions}, we obtain for
$k=0$ and $\mu\leq 0$
\begin{eqnarray}\label{mphiFinal}
\nonumber
\bar{m}_{\varphi,0}^2 &=& \mu_{\text{M}} (B-B_0)-2\mu\\
\nonumber
&& +\frac{\bar h_{\varphi,\Lambda}^2}{6\pi^2} {\Bigg [}\Lambda-\sqrt{\Lambda^2+\mu}
\left(1-\frac{\mu}{2\Lambda^2}\right)\\
&& +\frac{3}{2}\sqrt{-\mu} \,\,\text{arctan}
\left(\frac{\sqrt{\Lambda^2+\mu}}{\sqrt{-\mu}}\right){\Bigg ]}.
\end{eqnarray}
In the limit $\Lambda/\sqrt{-\mu}\to\infty$ this yields
\begin{equation}
\bar m_{\varphi,0}^2 = \mu_{\text{M}} (B-B_0)-2\mu+\frac{\bar{h}_{\varphi,\Lambda}^2\sqrt{-\mu}}{8\pi}.
\end{equation}
Together with Eq. \eqref{eqscatterinlengthandmageticfield}, we can deduce
\begin{equation}
a=-\frac{\bar h_{\varphi,\Lambda}^2}{8\pi \left( \bar m_{\varphi,0}^2 +2\mu 
-\frac{\bar{h}_{\varphi,\Lambda}^2\sqrt{-\mu}}{8\pi}\right)},
\end{equation}
which holds in the vacuum for all $\mu$. On the BEC side where $\bar
m_{\varphi,0}^2=0$ this yields
\begin{equation}
  a=\frac{1}{\sqrt{-\mu}\left(1+\frac{16 \pi}{\bar h_{\varphi,\Lambda}^2}\sqrt{-\mu}\right)}.
\label{ScattLength}
\end{equation}
The binding energy of the bosons is given by the difference between the
energy for a boson ${\bar{m}_\varphi^2}/{\bar{Z}_\varphi}$ and the energy
for two fermions $-2\mu$. On the BEC side, we can use $\bar{m}_{\varphi,0}^2=0$ and
obtain
\begin{equation}
\epsilon_{\text{M}}=\frac{\bar{m}_\varphi^2}{\bar{Z}_\varphi}+2\mu\Big|_{k\to0}=2\mu.
\label{eq:bindingenergy}
\end{equation}
From Eqs. \eqref{ScattLength} and \eqref{eq:bindingenergy} we find a relation
between the scattering length $a$ and the binding energy $\epsilon_\text{M}$
\begin{equation}
\frac{1}{a^2}=\frac{-\epsilon_\text{M}}{2}
+(-\epsilon_\text{M})^{3/2} \frac{4\sqrt{2}\pi}{\bar h_{\varphi,\Lambda}^2}
+(-\epsilon_\text{M})^2\frac{(8\pi)^2}{\bar h_{\varphi,\Lambda}^4}.
\label{eq:scatteringlengthandbindingenergy}
\end{equation}
In the broad resonance limit $\bar{h}_{\varphi,\Lambda}^2\to\infty$, this is
just the well-known relation between the scattering length $a$ and the binding
energy $\epsilon_{\text{M}}$ of a dimer
\begin{equation}
\epsilon_{\text{M}}=-\frac{2}{a^2}=-\frac{1}{M a^2}.
\label{eq:scatteringlengthbroad}
\end{equation}
The last two terms in Eq. \eqref{eq:scatteringlengthandbindingenergy} give
corrections to Eq. \eqref{eq:scatteringlengthbroad} for more narrow
resonances. Only in this case, the coupling $\bar h_{\varphi,\Lambda}$ acquires an independent physical meaning; for broad resonances, the only physical observable is the scattering length $a$ in Eq. \eqref{eq:hvarphiLambda}.

The solution of the two-body problem turns out to be exact as expected. In our
formalism, this is reflected by the fact that the two-body sector decouples
from the flow equations of the higher-order vertices: no higher-order
couplings such as $\lambda_\varphi$ enter the set of equations
(\ref{eq:vacuumflow}). Inspection of the full diagrammatic structure of our
truncation shows that there is a tadpole contribution $\sim \lambda_\varphi$
at arbitrary $n$ and $T$ which enters the flow equation for the bosonic
mass. However, the contribution of the corresponding diagram to the flow
vanishes in vacuum. Extending the truncation to even higher order vertices or
by including a boson-fermion vertex $\psi^\dagger \psi \varphi^\ast\varphi$
does not change the situation, since there are only two external lines in the
two-body problem and the flow equation involves only one-loop diagrams, such
that contributions from these vertices do not appear. We emphasize that this
argument is \emph{not} bound to a certain truncation.

\subsection{Universality}
Universality means that the macroscopic physics (on the length scales of the
order of the inter-particle distance) becomes independent of the details of
the microphysics (on the molecular scales) to a large extent. This is due to
the presence of fixed points in the renormalization flow. Approaching the
fixed point, the flow ``loses memory'' of most of the microphysics, except
for a few relevant parameters. We will show here how the broad resonance
universality emerges within the flow equations for the vacuum. The fixed point
structure remains similar for non-vanishing density and temperature, such that
our findings can easily be taken over to this more complex situation.

Let us consider the flow equations \eqref{eq:vacuumflow} for the renormalized
quantities ($m_\varphi^2=\bar m_\varphi^2/A_\varphi$, $\bar h_\varphi^2=h_\varphi^2/A_\varphi$, $S_\varphi=Z_\varphi/A_\varphi$) in the
regime where $k^2\gg -\mu$,
\begin{eqnarray}\label{ScalFlow}
\nonumber
  \partial_t m_\varphi^2 &=& \frac{k}{6\pi^2} h_\varphi^2 + \eta_{A_\varphi} m_\varphi^2, \\
\nonumber
\partial_t  S_\varphi  &=& -\frac{ 1}{6\pi^2 k} h_\varphi^2 + \eta_{A_\varphi} S_\varphi,\\
\nonumber
\partial_t  h_\varphi^2  &=& \eta_{A_\varphi} h_\varphi^2,\\
\eta_{A_\varphi} &=& -\frac{\partial_t A_\varphi}{A_\varphi}= \frac{1}{6\pi^2 k} h_\varphi^2.
\end{eqnarray}
In this scaling regime, the flow loses memory of all initial conditions
(except for the mass term which is a relevant parameter), provided that
the dimensionless combination $h_\varphi^2/k$ is attracted to a non-perturbative fixed point. The flow equation
\begin{equation}
  \partial_t \left(\frac{h_\varphi^2}{k}\right) =(-1+\eta_{A_\varphi})
  \frac{h_\varphi^2}{k}
\end{equation}
exhibits indeed an IR-attractive fixed point (scaling solution) given by $
\eta_{A_\varphi}=1$, $h_\varphi^2/k=6\pi^2$.  This fixed point is approached
rapidly provided the initial value of $h_\varphi^2(\Lambda)/\Lambda$ is large
enough. Memory of the precise value of $h_\varphi^2(\Lambda)/\Lambda$ is then
lost. The situation is similar for most other possible additional couplings
that we have not displayed here: The dimensionless renormalized couplings are
all attracted to fixed point values. For example, this also applies to the
four-boson coupling $\lambda_\varphi$ \cite{Diehl:2007XXX}.

A notable exception is the mass term $m_\varphi^2$. The dimensionless combination obeys
\begin{eqnarray}
\nonumber
\partial_t \left(\frac{m_\varphi^2}{k^2}\right) &=& \frac{1}{6\pi^2}\frac{h_\varphi^2}{k}-(2-\eta_{A_\varphi})\frac{m_\varphi^2}{k^2} =1-\frac{m_\varphi^2}{k^2},
\end{eqnarray}
where the fixed-point value for $h_\varphi^2/k$ has been used in the second
equation. The fixed point at $m_\varphi^2/k^2=1$ is unstable for the flow
towards the infrared. A small initial deviation from the fixed-point value
tends to grow as $k$ is lowered. The mass term is therefore a relevant
parameter. For $k\to 0$, $m_\varphi^2$ approaches a finite value, see
  Eq.~(\ref{eq:mvac}), and sets the scale for all dimensionful quantities on
  the BCS-side of the crossover.

We conclude that the precise initial values of all quantities except for the mass
term are unimportant. For definiteness, we choose $A_\varphi=Z_\varphi = 1$, $\lambda_\varphi = 0$.

Following the $k$ evolution further, on the BEC side for $\mu<0$ the flow
leaves the scaling regime near $k\sim \sqrt{-\mu}$. Thus a negative $\mu$ also
plays the role of a relevant parameter, similar to $m_\varphi^2$ for
$\mu=0$. The IR limit $k = 0$ is determined by a single scale set by the value
of $\mu$. This reflects the single relevant parameter which characterizes the
infrared physics of a broad resonance on the BEC side. We consider the flow of
$S_\varphi$ with $h_\varphi^2/k$ assuming its fixed point value,
\begin{eqnarray}\label{AFlow}
 \partial_t  S_\varphi  &=& (S_\varphi - 1)\eta_{A_\varphi}.
\end{eqnarray}
This equation is solved by $S_\varphi=Z_\varphi/A_\varphi = 1$ for $k \to 0$, corresponding to the dispersion relation for particles of mass $2M$.

To summarize, the flow starts at some microscopic scale $\Lambda$ with a
fine-tuned choice of $\bar m_{\varphi,\Lambda}^2>0$ given by Eq.
\eqref{eq:mvarphiLambda}. For broad resonances, the precise initial values of
all other parameters of our truncation ($S_\varphi$, $\lambda_\varphi$ ,
\dots) are unimportant as a consequence of universality -- the only relevant
parameter is the inverse scattering length.  In particular we see how a
molecular bound state appears on the BEC side of the crossover, independently
of the precise form of the microscopic physics which may be described by a
simple pointlike interaction.

\subsection{Dimer-Dimer Scattering}
\label{DimerDimer}

So far we have considered the sector of the theory up to order
$\varphi^\ast\psi\psi$, which is equivalent to the fermionic two-body problem
with pointlike interaction in the limit of broad resonances. Higher-order
couplings, in particular the four-boson coupling
$\lambda_\varphi(\varphi^\ast\varphi)^2$, do not couple to the two-body
sector. Nevertheless, a four-boson coupling emerges dynamically from the
renormalization group flow. 
In vacuum we have $\rho_0=0$ and $\lambda_\varphi$ is defined as $\lambda_\varphi=U^{\prime\prime}_k(0)$, cf. Eq. \eqref{2L}. The flow equation for $\lambda_\varphi$ can be found by taking the $\rho$-derivative of Eq. \eqref{eq:Flowu1}
\begin{eqnarray}\label{eq:17}
\nonumber
k \partial_k \lambda_\varphi &=& 2 \eta_{A_\varphi} U_k^{\prime\prime} - \frac{\sqrt{2} k^3}{3\pi^2 S_\varphi}\left(1-\frac{2}{d+2}\eta_{A_\varphi}\right)\\
\nonumber
&&\times 2 (U_k^{\prime\prime})^2 \left(s_{B,Q}^{(1,0)}+3 s_{B,Q}^{(0,1)}\right)+\frac{h_\varphi^4}{3\pi^2 k^3} s_{F,Q}^{(2)}\\
\nonumber
&=& 2\eta_{A_\varphi} \lambda_\varphi + \frac{\sqrt{2} k^5 \lambda_\varphi^2}{3\pi^2\, S_\varphi\, (m_\varphi^2+k^2)^2}(1-2\eta_{A_\varphi}/5)\\
&&-\frac{h_\varphi^4\,\theta(\mu+k^2)\,(\mu+k^2)^{3/2}}{4\pi^2k^6}.
\end{eqnarray} 
There are contributions from fermionic \emph{and} bosonic vacuum
fluctuations (cf. Fig. \ref{IntFlow}), but no contribution from higher $\rho$ derivatives of
$U$ as can be checked by applying the arguments from Sect.
\ref{sec:simplif}. The fermionic diagram generates a four-boson
coupling even for zero initial value. This coupling then feeds back
into the flow equation via the bosonic diagram.

First, we discuss the implications of broad-resonance universality for the
four-boson coupling and the ratio of the scattering length $a_{\text{M}}$ for
molecules and $a$ for fermions. In the scaling regime $k^2 \gg -\mu $ and
at large $\bar h_\varphi^2$, we use the scaling form Eq.  (\ref{ScalFlow})
implying $h_\varphi^2=6\pi^2 k$, $\eta_\varphi=1$, $S_\varphi=1$, $m_\varphi^2=k^2$. 
The dimensionless ratio $Q=\lambda_\varphi k^3/h_\varphi^4$ obeys the flow equation
\begin{equation}
 k \partial_k Q = 3 Q -\frac{1}{4\pi^2} + \frac{9 \sqrt{2}\pi^2}{5}\,  Q^2. \label{AZ31}
\end{equation}
The flow exhibits an infrared stable fixed point. Its value $Q_* \simeq 0.008$ corresponds to a
renormalized coupling scaling $\sim k^{-1}$,
\begin{equation}
  \lambda_\varphi =  \frac{36 \pi^4 Q_*}{k}. \label{AZ32}
\end{equation}
This can be compared to the effective four-fermion coupling in the
scaling regime, $\lambda_{\psi,\text{eff}}=-h_\varphi^2/m_\varphi^2 =- 6\pi^2/k$. Thus $\lambda_\varphi/\lambda_{\psi,\text{eff}}$ approaches a fixed point. The constant ratio
between these two quantities in the scaling regime of the flow is the origin of the universal ratio between the scattering length for molecules and atoms.

Having settled the universality of the scattering length ratio, we now
address its numerical value. The scattering lengths are related to the
corresponding couplings by the relation (cf. \cite{Diehl:2007XXX})
\begin{eqnarray}
  \frac{a_{\text{M}}}{a}=2\,
  \frac{\lambda_\varphi}{\lambda_{\psi,\text{eff}}},  
  \quad \lambda_{\psi,\text{eff}}= 8\pi c = 
  \frac{8\pi}{\sqrt{-{\mu}}} .
\end{eqnarray}
The last equality is obtained in the molecule phase for $\mu<0$ and
$k=0$, using Eq. (\ref{mphiFinal}) together with the constraint (\ref{VacCond}). 
Omitting the molecule fluctuations, a direct integration of Eq.~\eqref{eq:17}
yields $\lambda_\varphi=8\pi/\sqrt{-\mu}$ and therefore
$a_{\text{M}}/a=2$. Inclusion of the molecule fluctuations lowers this
ratio. We stress that the vacuum value for $\lambda_\varphi/\lambda_{\psi,\text{eff}}$ has to be computed for $k\to 0$. It therefore differs somewhat from the ratio in the scaling regime (Eq. \eqref{AZ32}) since the final flow for $k^2\lesssim -\mu$ has to be taken into account. The result is weakly cutoff dependent. With our truncation and choice of cutoff one finds $a_{\text{M}}/a =0.718$.

The ratio $a_M/a$ has been computed by other methods. Diagrammatic approaches give $a_{\text{M}}/a =0.75(4)$
\cite{AAAAStrinati}, whereas the solution of the 4-body Schr\"{o}dinger
equation yields $a_{\text{M}}/a =0.6$ \cite{Petrov04}, confirmed in QMC
simulations \cite{Giorgini04} and with diagrammatic techniques
\cite{Kagan05}. Our calculation can be improved by extending the truncation to
include a boson-fermion vertex $\lambda_{\varphi\psi}$ which describes the
scattering of a dimer off a fermion \cite{DSK08}. Inspection of the diagrammatic structure
shows that this vertex indeed couples into the flow equation for
$\lambda_\varphi$.  Moreover, this coupling is important for a precision
estimate of the dynamically generated atom-dimer scattering length.

\section{Particle-hole fluctuations}
\label{sec:phfluct}

In the previous section we were concerned with the flow equations in
the vacuum limit where both the density and the temperature vanish,
$n=T=0$. It is one of the main advantages of our method that we can treat this
limit as well as the many-body problem in thermal equilibrium with nonzero
density and temperature in a unified approach. On large scales $k^2\gg T$,
$k^2\gg n^{2/3}$ the flow equations are basically the same as in the vacuum
case. The qualitative features and the fixed points are the same as discussed in the
previous section. However, for scales of the order of the inverse particle
distance $k\approx n^{1/3}$ the flow equations are modified by many-body
effects. We have seen in
Sec. \ref{sec:SuperfluidityPhaseTransition} how the fermionic
fluctuations lead to a nonvanishing order parameter $\rho_0>0$ for
$k<k_\text{SSB}$ connected to local order, and to superfluidity if $\rho_0$
remains positive for $k=0$.  

\begin{figure}[ht]
\centering
\includegraphics[width=0.35\textwidth]{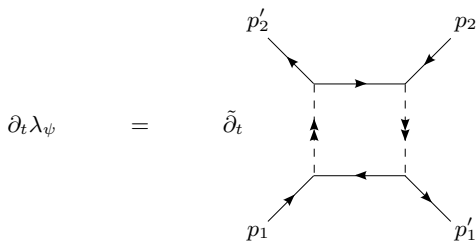}
\caption{Box diagram for the flow of the four-fermion interaction. Arrows
  denote the flow of the particle number.}
\label{fig:boxes}
\end{figure}

Other interesting many-body effects are the screening
of the four-boson interaction $\lambda_\varphi$ on the BEC side of the
crossover or the effect of particle-hole fluctuations on several thermodynamic
observables on the BCS side.  The first effect is well captured in our present
truncation. It manifests itself by additional terms in the flow equation for
$\lambda_\varphi$ which are not present in the vacuum limit. These bosonic
fluctuations lead to a decrease of the four-boson interaction
$\lambda_\varphi(k=0)$. This is very similar to the screening effect in a
system of non-relativistic bosons with pointlike interaction
\cite{FloerchingerWetterich, FWBoseD2}. The effect of particle-hole fluctuations is
 more involved. The truncation discussed in Sect. \ref{sec:trunc} is not
yet sufficient to capture it. We next discuss the extensions necessary to include this effect following Ref. \cite{FSDW08}.
Within the approach discussed in Sect. \ref{Method}, the effect of
particle-hole fluctuations generates a four fermion vertex $\lambda_\psi$ for $k<\Lambda$, even if we start with $\lambda_{\psi,\Lambda}=0$ due to bosonization of the original pointlike interaction. The total effective four-fermion interaction is composed of a
contribution from the exchange of a boson and of $\lambda_\psi$ generated by the renormalization flow from the box-diagrams involving
the fermions and the bosons as displayed in Fig. \ref{fig:boxes}, 
\begin{equation}
\lambda_{\psi,\text{eff}}=-\frac{\bar h_\varphi^2}{\bar P_\varphi}+\lambda_\psi.
\end{equation}
The vertex $\lambda_\psi$ is generated only in medium, i.e. at finite temperature and density while the corresponding diagram vanishes in the vacuum.

If we contract the dashed boson exchange lines in Fig. \ref{fig:boxes} to a point it is easy to see that it describes the interaction of two incoming particles (with momenta $p_1$, $p_2$ and similar for the outgoing channel) with a (virtual) particle-hole pair. Using the formalism of rebosonization developed in \cite{Gies:2001nw} it is
possible to adjust the boson exchange in a scale-dependent manner such
that the effect from particle-hole fluctuations is included in the boson exchange
description. This method involves scale-dependent fields where the boson is an
effective degree of freedom. Technically, the use of scale-dependent fields
leads to a modified flow equation for the Yukawa coupling $h_\varphi$. For
details we refer to Ref. \cite{FSDW08}.

Most prominently, the effect of particle-hole fluctuations is seen in the
critical temperature $T_{\text{c}}$.  For non-relativistic fermions the task
of a precise computation of $T_{\text{c}}$ is difficult, the basic reason
being that the phase transition itself is a non-perturbative phenomenon,
linked to an effective four-fermion interaction $\lambda_{\psi,\text{eff}}$
growing to large values. Indeed, the Thouless criterion states that a phase
transition is connected with a diverging effective interaction between
fermions $\lambda_{\psi,\text{eff}}\to\infty$. In order to locate
$T_{\text{c}}$ precisely, one has to follow the scale-dependence of
$\lambda_{\psi,\text{eff}}$ with sufficient precision. The problem arises from
the substantial momentum dependence of the fluctuation contributions to the
effective four-fermion vertex. At weak coupling, the dominant deviation from
the BCS result for $T_{\text{c}}$ -- which is obtained by neglecting the
momentum dependence of the four-fermion vertex completely -- comes from the
in-medium screening due to the particle-hole fluctuations. It was first shown
by Gorkov and Melik-Barkhudarov \cite{Gorkov} (see also \cite{Heiselberg}) in a perturbative setting that
this effect leads to a decrease of the critical temperature in the BCS regime
by a factor $\sim 2.2$ compared to the result obtained in the original BCS
theory.

\begin{figure}[ht]
\centering
\includegraphics[width=0.45\textwidth]{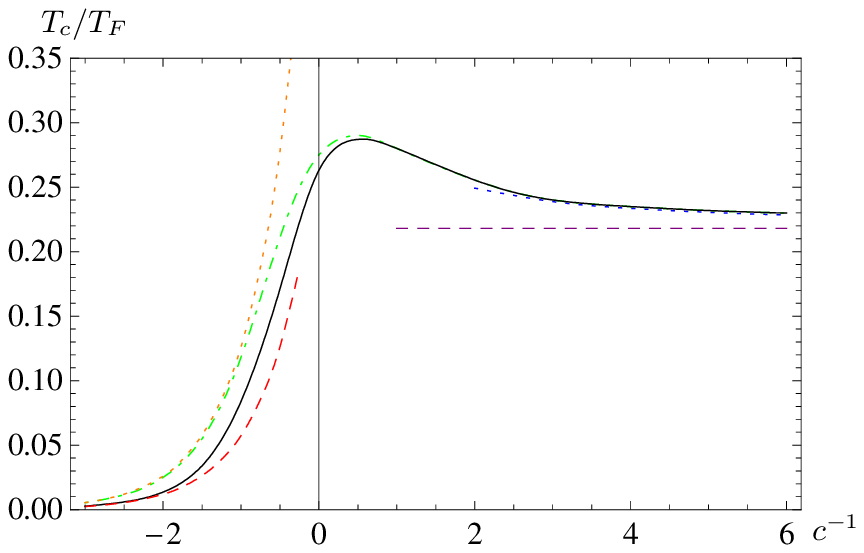}
\caption{(Color online) Dimensionless critical temperature
  $T_{\text{c}}/T_{\text{F}}$ as a function of the inverse concentration
  $c^{-1}=(a k_\mathrm{F})^{-1}$. This plot is taken from Ref. \cite{FSDW08}. The black
  solid line includes the effect of particle-hole fluctuations. The result
  obtained when particle-hole fluctuations are neglected is also shown
  (dot-dashed line). For comparison, we plot the BCS result without (left
  dotted line) and with Gorkov's correction (left dashed). On the BEC side
  with $c^{-1}>1$ we show the critical temperature for a gas of free bosonic
  molecules (horizontal dashed line) and a fit to the shift in $T_{\text{c}}$
  for interacting bosons, $\Delta T_{\text{c}}\sim c$ (dotted line on the
  right).}
\label{fig:tcrit}
\end{figure}

We show the numerical result of Ref. \cite{FSDW08} for $T_{\text{c}}/T_{\text{F}}$ in Fig. \ref{fig:tcrit}, together with the critical temperature obtained in the truncation (\ref{eq:renormalizedtruncation}).  
The results from the extended truncation used in \cite{FSDW08} are
  depicted be the solid line in Fig.~\ref{fig:tcrit}. This result agrees with
  the prediction by Gorkov and Melik-Barkhudarov in the regime with small
  negative scattering length (long-dashed line in Fig. \ref{fig:tcrit}). By
  contrast, the simpler truncation described in Sec. \ref{Method} of this paper (dot-dashed line)
  reproduces the original BCS result (dotted line in
  Fig. \ref{fig:tcrit}). Both our approximations yield the same result for
  $c^{-1}\gtrsim 0.5$ where the effect of particle-hole fluctuations
  disappears. This is expected since the chemical potential is negative here,
  $\mu<0$, and there is no Fermi surface any more. In the BEC limit for very
small positive scattering length we find that our result approaches the
critical temperature of a free Bose gas
\begin{equation}
\frac{T_{\text{c,BEC}}}{T_{\text{F}}}\approx 0.218.
\end{equation} 
For $c\to 0_+$ this value is approached in the form
\begin{equation}
\frac{T_{\text{c}}-T_{\text{c,BEC}}}{T_{\text{c,BEC}}}=\kappa a_{\text{M}} n_{\text{M}}^{1/3}=\kappa \frac{a_{\text{M}}}{a}\frac{c}{(6\pi^2)^{1/3}}.
\label{eq:shiftTcBEC}
\end{equation}
Here, $n_{\text{M}}=n/2$ is the density of molecules and $a_{\text{M}}$ is the scattering length
between them. For the ratio $a_{\text{M}}/a$ we use our result $a_{\text{M}}/a=0.718$ obtained
from solving the flow equations in vacuum, see sect. \ref{Vacuum}, since for tightly bound molecules this is the relevant microscopic interaction parameter. For the
coefficients determining the shift in $T_{\text{c}}$ compared to the free Bose
gas we find $\kappa=1.55$. Given the simplicity of our approach, it is
remarkable that our result is near the value $\kappa\simeq 1.3$ found in an effective three dimensional ``classical'' bosonic theory \cite{Baymetal} with lattice simulations \cite{Arnold:2001mu,kashurnikov:2001} and with the
functional RG using elaborate momentum-dependent truncations
\cite{Blaizot:2006vr}. A direct quantitative comparison is, however, inhibited
by the fact that our result may receive contributions from the
residual fermionic modes.

At the unitarity point $c^{-1}=0$ we find $T_{\text{c}}/T_{\text{F}}=0.264$
with the approximation used in \cite{FSDW08}. The simpler truncation used in Sect \ref{Method}
gives $T_{\text{c}}/T_{\text{F}}=0.276$. This should be compared with results
from Quantum Monte Carlo simulations: $T_{\text{c}}/T_{\text{F}} = 0.15$
\cite{MonteCarloBuro,MonteCarloBulgac} and $T_{\text{c}}/T_{\text{F}} = 0.245$
\cite{MonteCarloAkki}. The measurement by L.~Luo \textit{et al.}
\cite{Luo2007} in an optical trap gives $T_{\text{c}}/T_{\text{F}} = 0.29
(+0.03/-0.02)$, which is a result based on the study of the specific heat of
the system. Further theoretical predictions based on a variety of
  methods span a whole range of values: $T_{\text{c}}/T_{\text{F}}=0.249$
  ($\epsilon$ expansion \cite{Nishida:2006br}),
  $T_{\text{c}}/T_{\text{F}}=0.183$ (Borel-Pad\'{e} approximation
  \cite{Nishida:2006eu}), $T_{\text{c}}/T_{\text{F}}=0.136$ ($1/N$ expansion
  \cite{Nikolic:2007zz}), $T_{\text{c}}/T_{\text{F}}=0.160$ (2PI techniques
  \cite{Haussmann:2007zz}). This demonstrates that quantitative precision
requires a control of the systematic errors in a given approximation
scheme. A particular important point concerns the quantitative accuracy in the determination of the density. We will discuss this issue in Sect. \ref{sec:Density}. If we replace our determination of the density by the ``free particle density'' we obtain (in the truncation of Sect. \ref{Method})  $T_{\mathrm{c}}/T_{\mathrm{F}}=0.171$.

\section{Universal long-range physics close to the phase transition}
\label{sec:ClosePt}

As the phase transition is approached, one expects to find universal critical
behaviour if the temperature is close enough to the critical temperature
$T_{\mathrm{c}}$. This feature can be seen directly from the nonperturbative
flow equations. The universal behaviour arises from long-range bosonic
fluctuations with momenta below $T^{1/2}$ and larger than the inverse
correlation length $\xi^{-1}$, and requires $\xi T_{\mathrm{c}}^{1/2} \gg
1$. The temperature acts as an effective infrared cutoff for the influence of
the fermion fluctuations on the long-range physics. Correspondingly, for
$k/\sqrt{T_{\mathrm{c}}}\ll 1$ the fermion contribution to the flow becomes
small and can be neglected. This can be inferred directly from the dependence
of the fermionic threshold functions on $k/\sqrt{T}$, as we have discussed
after Eq. \eqref{eq:Floweqhightemperature}. On the other hand, for
$k/\sqrt{T}\ll 1$ the bosonic threshold functions are transmuted to the ones
of a classical statistical system for bosons with U(1) symmetry. The infrared
flow is governed by a Wilson-Fisher fixed point for the $N=2$ universality
class, where $N=2$ corresponds to the two real scalar fields on which global
$SO(2)$-transformations act linearly. Following the flow equations to $k\to0$
leaves no doubt that this universality class governs the critical exponents
and amplitude ratios for $T$ sufficiently close to $T_{\mathrm{c}}$. For more
elaborate truncations, the non-perturbative flow equations yield precise values
for these quantities \cite{Berges:2000ew, Blaizot:2006vr}.

\begin{figure}
\begin{minipage}{\linewidth}
\begin{center}
\setlength{\unitlength}{1mm}
\begin{picture}(82,57)
\put (0,0){
    \makebox(80,56){
\begin{picture}(80,56)
      \put(0,0){\epsfxsize80mm \epsffile{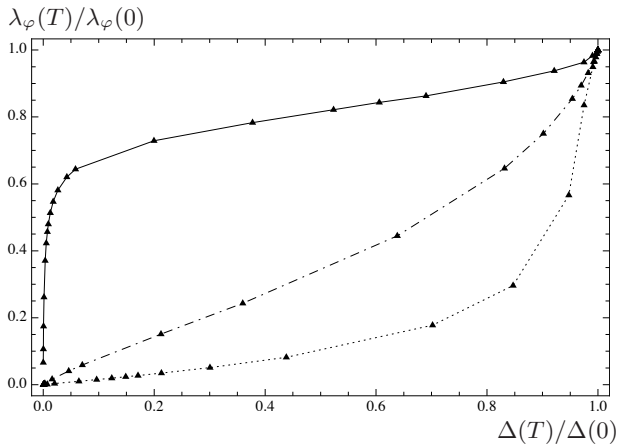}}
      \put(65,-3){$\Delta(T)/\Delta(0)$}
      \put(0,52){$\lambda_\varphi(T)/\lambda_\varphi(0)$}
\end{picture}
      }}
   \end{picture}
\end{center}
\vspace*{-1.25ex} \caption{Boson self-interaction $\lambda_\varphi /
  \lambda_\varphi (T=0)$ as a function of the gap parameter $\Delta (T)/
  \Delta(T=0)$. The corresponding temperature range is $0\leq T \leq
  T_{\text{c}}$, where $T_{\text{c}}$ corresponds to the origin of the
  plot. Solid line: BCS regime, $c^{-1}= -2$. Dot-dashed line: crossover
  regime, $c^{-1} = 0$. Dotted line: BEC regime, $c^{-1} =4$. }
\label{PhaseTransition2}
\end{minipage}
\end{figure}

It is not our aim here to reproduce the precise universal quantities by an
extension of our truncation. We are rather interested in another question: how
extended is the temperature range for the universal behavior? Furthermore, we
want to compute the non-universal aspects of the phase transition as
well. They will depend, of course, on the precise location of the system in
the BCS-BEC crossover. As an example, we investigate the temperature
dependence of the boson self-interaction $\lambda_\varphi$. Close to the
critical temperature, $\lambda_\varphi$ is known to vanish
($\lambda_\varphi^2/m_\varphi^2$ goes to a constant in the symmetric
phase). Since also the gap $\Delta$ vanishes for $T\to T_{\mathrm{c}}$, we may
take the size of the gap $\Delta(T)$ as a measure of the distance from the
critical temperature. In Fig. \ref{PhaseTransition2} we plot the ratio
$\lambda_\varphi(T)/\lambda_\varphi(T=0)$ as a function of
$\Delta(T)/\Delta(T=0)$.
The plots correspond to the BCS (solid; $c^{-1}= -2$), crossover (dashed
dotted; $c^{-1} = 0$) and BEC (dotted; $c^{-1} =4$) regimes. In the BEC and
unitarity regime, we observe an almost linear increase of $\lambda_\varphi$
with $\Delta$ -- neglecting effects of a small anomalous dimension, the
universal scaling relations indeed suggest $\lambda_\varphi/\Delta\to
\text{const}$. The linear increase of $\lambda_\varphi$ with $\Delta$ extends
within a substantial range in temperature. On the BCS side, in contrast, the
universal critical region strongly shrinks.

\begin{figure}
\begin{minipage}{\linewidth}
\begin{center}
\setlength{\unitlength}{1mm}
\begin{picture}(82,57)
\put (0,0){
    \makebox(80,56){
\begin{picture}(80,56)
      \put(0,0){\epsfxsize80mm \epsffile{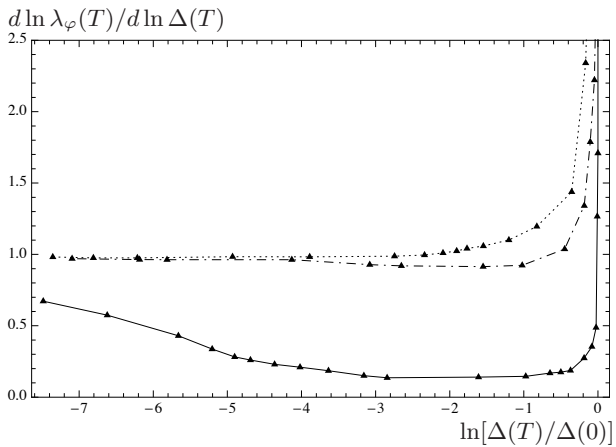}}
      \put(60,-3){$\ln[\Delta(T)/\Delta(0)]$}
      \put(0,52){$d \ln\lambda_\varphi(T)/d\ln\Delta(T)$}
\end{picture}
      }}
   \end{picture}
\end{center}
\vspace*{-1.25ex} \caption{Logarithmic slope of $\lambda_\varphi$,
  given by $d \ln \lambda_\varphi / d \ln \Delta$, as a function of the gap
  $\ln [\Delta (T) /\Delta (T=0) ]$. Here $T_{\text{c}}$ is approached for
  large negative values of $\ln \Delta$. Solid line: BCS regime, $c^{-1}=
  -2$. Dot-dashed line: crossover regime, $c^{-1} = 0$. Dotted line: BEC regime,
  $c^{-1} =4$.}
\label{PhaseTransition3}
\end{minipage}
\end{figure}

In Fig. \ref{PhaseTransition3}, we show the slope of $\lambda_\varphi/\Delta$
on a logarithmic scale.
For the BEC and unitarity regime, we can numerically read off the slope,
$\lambda_\varphi\sim \Delta^{\zeta}$, $\zeta=0.98$.  For the unitarity regime
($c^{-1}=0$, dashed dotted line), the universal behavior seems to be reached
very quickly, typically already for $\Delta(T)\approx \Delta(0)/2$. This is
related to the presence of a strong coupling which induces a fast approach to
the infrared fixed point. The universal region is smaller on the BEC side
($c^{-1}=4$, dotted line). In contrast, the approach to the universal fixed
point is very slow in the BCS regime ($c^{-1}=-2$, solid line). In
consequence, the temperature interval around $T_{\mathrm{c}}$ for which
universal behavior occurs is found to be very narrow.

\section{Atom density}
\label{sec:Density}

Since we use the grand canonical formalism, the particle number $n$ is fixed
indirectly by the chemical potential $\mu$ which is a parameter of our
microscopic model in Eq.\ \eqref{eq:action} similar as the temperature $T$ or
the detuning $\nu$. One of the dimensionful parameters can be used to set the
scale of the problem, such that actual computations only determine
dimensionless combinations of observables as functions of dimensionless
combinations of parameters. To compare our results to experiment or other
methods one might in principle consider dimensionless quantities that involve
$\mu$, for example $T_{\mathrm{c}}/\mu$. For example, with the truncation
described in Sect.\ \ref{Method}, we find at unitarity the ratio $T_{\mathrm{c}}/\mu=0.44$
while the improved truncation in Ref.\ \cite{FSDW08} yields
$T_{\mathrm{c}}/\mu=0.39$. This may be compared with the results from
Monte-Carlo calculations $T_{\mathrm{c}}/\mu=0.31$ \cite{MonteCarloBuro},
$T_{\mathrm{c}}/\mu=0.35$ \cite{MonteCarloBulgac}, from $1/N$-expansion
$T_{\mathrm{c}}/\mu=0.232$ \cite{Nikolic:2007zz} or from 2-PI methods
$T_{\mathrm{c}}/\mu=0.41$ \cite{Haussmann:2007zz}.

However, the rescaling with respect to $\mu$ is not optimal for a comparison to experiments. In contrast to the particle density $n$, the chemical potential $\mu$ is not directly accessible experimentally. Dimensionless ratios that directly involve the particle number such as $T_{\mathrm{c}}/T_{\mathrm{F}}$ with $T_{\mathrm{F}}=(3\pi^2 n)^{2/3}$ can be measured more directly than $T_{\mathrm{c}}/\mu$. 

For strongly interacting particles the dependence $n(\mu)$ is non-trivial to
obtain. This is in contrast to weak interactions where $n(\mu)$ can be
estimated by the corresponding formula for a non-interacting gas. Part of the
uncertainity in the prediction of a dimensionless ratio that involves the
density arises therefore from the determination of the function $n(\mu)$. In
this paper, we use a flow equation for a scale-dependent generalization of the
density,
\begin{equation}
\partial_k n_k = -\partial_k \frac{\partial}{\partial \mu} U_k,
\end{equation}
in order to calculate the density $n=n_{k}|_{k=0}$. The $\mu$ dependence of
$U_k$ is approximated by an expansion in $\rho$ and $\mu$ as shown in
Eq.\ \eqref{2L}. For the determination of $n$, we need only the terms linear in
$\delta \mu$ with $\mu=\mu_0+\delta \mu$. Besides the term linear in $\delta
\mu$ and $\rho-\rho_0$, one might add another term which is quadratic in
$\rho-\rho_0$. We have checked numerically that the inclusion of such a term
has only a very small influence on the quantitative determination of $n$ and
neglect it for our numerical results.

In this work, we neglect a possible renormalization of the fermion propagator
$G_\psi$. This implies also that the dependence on $\mu$ is the same for all
cutoff scales $k$. In particular, the effect of fluctuations on the size of
the Fermi sphere is not taken into account such that it is of radius
$\mu^{1/2}$ an all scales. Beyond our approximation, we expect that the effect
of a $k$-dependent Fermi sphere changes somewhat the dependence of the density
$n$ on the chemical potential $\mu$.

\begin{figure}
\begin{minipage}{\linewidth}
\begin{center}
\setlength{\unitlength}{1mm}
\begin{picture}(82,57)
\put (0,0){
    \makebox(80,56){
\begin{picture}(80,56)
      \put(0,0){\epsfxsize80mm \epsffile{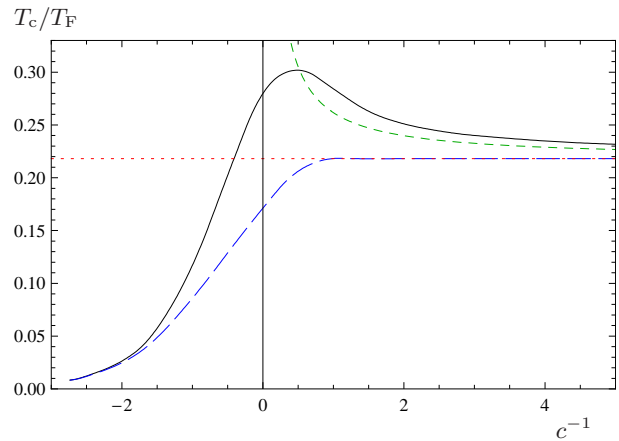}}
      \put(72,-3){$c^{-1}$}
      \put(0,52){$T_{\mathrm{c}}/T_{\mathrm{F}}$}
\end{picture}
      }}
   \end{picture}
\end{center}
\vspace*{-1.25ex} \caption{(Color online) Dimensionless critical temperature
  $T_{\text{c}}/T_{\text{F}}$ as a function of the inverse concentration
  $c^{-1}=(a k_\mathrm{F})^{-1}$. We show the result obtained in the truncation descibed in Sec. \ref{Method} with the flow equation for the density (solid line).  In addition we show the result obtained using the simpler (naive) estimate for the density in Eq. \eqref{eq:nnaive} (long dashed line). The dotted line gives the critical temperature of a free BEC while the short-dashed line is the expected correction for an interacting Bose gas in Eq.\ \eqref{eq:shiftTcBEC}, with the Monte-Carlo result $\kappa=1.3$ and with $a_M/a=0.6$ \cite{Petrov04}.}
\label{fig:density}
\end{minipage}
\end{figure}

To illustrate the importance of the density determination, we compare our result for $T_{\mathrm{c}}/T_{\mathrm{F}}$ with a calculation where a much simpler (and naive) method is used to estimate the density. It is motivated physically by assuming that we have simply coexisting pointlike atoms and dimers, which are coupled via a common chemical potential  \cite{Diehl:2005ae} -- in this picture, strong correlation effects are completely omitted for the density determination, and we can get an impression of the sensitivity of our results with respect to such correlation effects. At the critical temperature the superfluid and condensate densities vanish and no occupation number density for a single momentum mode is expected to be of the order $n$. In addition, the energy gap for the bosons just vanishes, $m_\varphi^2=0$. Neglecting the effect of interactions, one could use a naive estimate for the density at $T=T_{\mathrm{c}}$,
\begin{equation}
n_\text{naive}=\int \frac{d^3 p}{(2\pi)^3}\left\{\frac{2}{e^{(\vec p^2-\mu)/T_{\mathrm{c}}}+1}+\frac{2}{e^{\vec p^2/(2T_{\mathrm{c}})}-1}\right\}.
\label{eq:nnaive}
\end{equation}
The first term gives the contribution of the fermions where the factor $2$
counts the spin-states. As stated above, the second contribution comes from the bosons and the
factor $2$ reflects now that every boson consists of two fermions. Our flow equation method yields a result for $n$ that is roughly a
factor $1/2$ smaller than $n_\text{naive}$. Correspondingly, the ratio
$T_{\mathrm{c}}/T_{\mathrm{F}}$ is larger. For a comparison of both methods, we
plot $T_{\mathrm{c}}/T_{\mathrm{F}}$ as a function of $c^{-1}$ in
Fig. \ref{fig:density}.  At the unitarity point $c^{-1}=0$, the formula in
Eq. \eqref{eq:nnaive} leads to $T_{\mathrm{c}}/T_{\mathrm{F}}=0.171$.

As a side remark, we note that the approximation made to obtain Eq.\ \eqref{eq:nnaive} becomes valid for extremely narrow resonances with vanishing Yukawa interaction $h_\varphi\to0$ \cite{Diehl:2005ae}. Of course, the strongly interacting problem we are interested in here is the opposite limit of broad resonances $h_\varphi\to\infty$ -- here, we only try to estimate the uncertainty emerging from the neglect of correlation effects on the density. 

\begin{figure}
\begin{minipage}{\linewidth}
\begin{center}
\setlength{\unitlength}{1mm}
\begin{picture}(82,57)
\put (0,0){
    \makebox(80,56){
\begin{picture}(80,56)
      \put(0,0){\epsfxsize80mm \epsffile{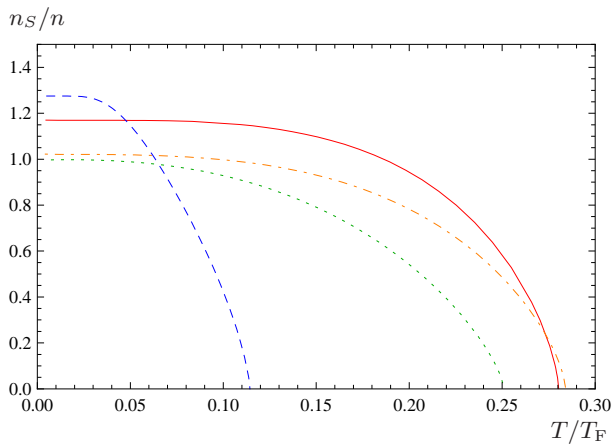}}
      \put(72,-3){$T/T_{\mathrm{F}}$}
      \put(0,52){$n_S/n$}
\end{picture}
      }}
   \end{picture}
\end{center}
\vspace*{-1.25ex} \caption{Superfluid fraction $n_{\mathrm{S}}/n=2\rho_0/n$ as
  a function of temperature $T/T_{\mathrm{F}}$. We show the curves obtained on
  the BCS side with $c^{-1}=-1$ (dashed line), in the unitarity regime with
  $c^{-1}=0$ (solid line) and on the BEC side with $c^{-1}=1$ (dashed-dotted
  line) and $c^{-1}=2$ (dotted line). While the superfluid fraction approaches
  $1$ for small temperatures on the BEC side, this is not the case on the BSC
  side and in the unitarity regime. This feature is an artefact of our present
  truncation.}
\label{fig:SD}
\end{minipage}
\end{figure}

At the phase transition we have two independent (dimensionless) ratios
$T_{\mathrm{c}}/T_{\mathrm{F}}$ and $\mu/T_{\mathrm{F}}$. For the unitarity
limit, $c^{-1}=0$ we display these ratios for different approximations and
compare them to Monte-Carlo results in table \ref{tab:results}. The lack of
agreement may indicate that the truncation should be improved further, and we
discuss various directions in the conclusions. Another insufficiency of the
present truncation becomes visible in the behaviour of the superfluid density
for $T\to 0$, as shown in Fig.\ \ref{fig:SD}. The ratio $n_{\mathrm{S}}/n$ should remain
smaller than one, which is not obeyed by our current truncation on the BCS side. At zero temperature Galilean invariance implies $n=n_S$ throughout the whole BCS-BEC crossover. We expect that the problem can be solved by including the $\rho$ dependence of the wavefunction renormalization $A_\varphi$ where the dependence on $\rho$ was neglected in our truncation.
\begin{table}
\begin{tabular}{lccc}
\hline\hline
 & $\,\,\,T_{\mathrm{c}}/T_{\mathrm{F}}\,\,\,$ & $\,\,\,\mu/T_{\mathrm{F}}\,\,\,$ & $\,\,\,T_{\mathrm{c}}/\mu\,\,\,$ \\\hline
Truncation Sect. \ref{Method} & 0.276 & 0.63 & 0.44 \\\hline
Truncation including & 0.245 & 0.63 & 0.39 \\
particle-hole fluc. \cite{FSDW08} & &  & \\\hline
Density according & 0.171 & 0.39 & 0.44 \\
to Eq.\ \eqref{eq:nnaive} & & & \\\hline
Monte-Carlo \cite{MonteCarloBuro} & 0.152(7)  & 0.493(17) & 0.31 \\\hline
Monte-Carlo \cite{MonteCarloBulgac} & 0.15(1) & 0.43(1) & 0.35\\\hline\hline
\end{tabular}
\caption{Critical temperature $T_{\mathrm{c}}$, critical density $n_c$ (expressed through the Fermi temperature $T_{\mathrm{F}}=(3\pi^2 n_c)^{2/3}$), and chemical potential $\mu$ in the ``unitarity limit'' of infinite scattering length $c^{-1}=0$.}
\label{tab:results}
\end{table}

\section{Conclusion}
\label{sec:Conclusions}

In this paper, we have given a rather detailed account of the use of the
flowing action for ultracold fermionic atoms. Our method allows for a unified
approach for all temperatures and densities -- including the vacuum for
$T=n=0$ -- as well as arbitrary scattering length. The overall picture of the
phase transition to the low-temperature superfluid phase and the BCS-BEC
crossover as a function of the scattering length or concentration $c=a\kF$ is
rather satisfactory. In particular, the dependence of the critical temperature
$T_{\mathrm{c}}$ on the scattering length yields the effects of interactions
between composite molecules on the BEC side, as well as the correct behaviour
on the BCS side including the effect of particle-hole fluctuations. The
largest uncertainties occur in the ``unitarity limit'' $c^{-1}=0$ of very
large scattering length. In this region in parameter space, the fluctuations
are strongest and possible deviations of the effective action from our rather
simple truncation are expected to be largest. As far as the critical
temperature for $c^{-1}=0$ is concerned, the comparison with the Monte-Carlo
results in table \ref{tab:results} may give an idea of the uncertainties.

The phase diagram can be computed from the dependence of the effective boson
potential $U(\varphi)$ on the temperature and density or $T$ and $\mu$. The
exact flow equation \eqref{2D} for the flowing potential $U_k(\varphi)$
involves the exact propagators of fermions and composite bosons in the
presence of a constant background field $\varphi$ -- the condensate
corresponds to the value of $\varphi_0$ for which the minimum of $U(\varphi)$
occurs. The present truncation (Eqs.\ \eqref{2E}, \eqref{2I}) of the
propagators remains rather crude and we expect a considerable quantitative
improvement for truncations with a more detailed resolution of the dependence
of the propagators on momentum and frequency, as well as their dependence on
$\varphi$. On the other hand, increased precision also needs an increased
accuracy in the parameters used in a given truncation of the propagators. They
are determined by exact flow equations for the inverse propagators which
involve, in turn, the cubic and quartic vertices. In this respect the accuracy
may be enhanced by a more complicated truncation of the momentum and frequency
dependence of the effective four-fermion interaction, beyond the
molecule-exchange channel on which we have concentrated in this paper. In this
respect, methods of rebosonization which closely keep track of the relation
between the composite field $\varphi$ and its fermionic constituents
\cite{FWCompositeoperators} could lead to an impoved treatment of
particle-hole fluctuations for large scattering length.

The effort of improving the truncation leads not only to an increased accuracy
of the phase diagram. Simultaneously, many detailed features of the
propagators become available. We have demonstrated this briefly by our
computation of the correlation length. Furthermore, many thermodynamic
quantities can be computed from the dependence of $U$ on $T$ and $\mu$, as
specific heat, compressibility, or sound velocities. This has been
demonstrated for bosonic atoms in \cite{FWThermod} and the same methods could
be taken over to the BEC-BCS crossover system. Further generalizations concern
systems with three or more species of fermions \cite{FSMW} or unequal
abundances of different fermion species \cite{Gubbels:2008zz,Ku:2008vk}. We hope that
the generality and flexibility of our treatment will lead to a unified
treatment of many interesting features encountered in these systems, and
simultaneously permit quantitative accuracy of the theoretical estimates, as
needed for precision measurements.

\begin{acknowledgments}

This work has been supported by the DFG research unit FOR 723.
H.G.~acknowledges support by the DFG under contract Gi 328/1-4 (Emmy-Noether
program) and Gi 328/5-1 (Heisenberg-Program).

\end{acknowledgments}

\appendix

\section{Units}

In this work, our units are chosen such that $\hbar=k_{\text B}=2M=1$, where $M$ is the mass of the fermionic atoms. Setting $\hbar=1$ implies that space coordinates and inverse momenta are measured in
terms of the same dimensionful scale. The choice $k_{\text B}=1$ similarly
establishes an equivalent measurement of energy and temperature in terms of
the same scale. Setting $2M=1$ finally relates the measurement scales of
energy and temperature on the one hand to squared momenta or inverse squared
coordinates on the other hand. (The meaning of the latter choice is similar to
setting the velocity of light $c=1$ in relativistic theories). Only one unit remains free, say the length $l$, with dimensions $[t]=l^2$, $[\vec q]=l^{-1}$, $[T]=l^{-2}$, $[\mu]=l^{-2}$, and $[n]=l^{-3}$.

In practice, these units are obtained by rescaling the coordinates $x/\hbar\to
x$ for $\hbar=1$, the temperature $k_{\text{B}}T\to T$ for $k_{\text B}=1$, and
the time $\tau/2M \to \tau$ for $2M=1$. The last step goes along with
rescaling also all interaction terms of the action: $2M S_{\text{int}}\to
S_{\text{int}}$. All dimensionful quantities are then measured in terms of
only one dimensionful scale which we set in terms of the density $n$ of the
system or, equivalently, in terms of the Fermi momentum
\begin{equation}
\kF=(3\pi^2 n)^{1/3}. 
\end{equation}
Typical values of the Fermi momentum occurring in trapped Fermi-gas
experiments are of the order $\kF\sim \mathcal O( 1\dots 10\mathrm{eV}/c)$ in
units of electron Volt/$c$ (The light velocity $c$ occurs here to convert the
energy unit eV into a momentum). With our choice of units, $\kF=1$eV/$c$
corresponds to a density $n\simeq4.4\times 10^{12}\mathrm{cm}^{-3}$ for which
we have used that 1eV/$c\simeq 5.07\times 10^6\mathrm{m}^{-1}$ in our units
with $\hbar=1$. A typical length scale such as the Bohr radius
$a_{\text{B}}=5.29177\times 10^{-11}\mathrm{m}$ then reads in inverse momentum
units $a_{\text{B}}=2.6817\times 10^{-4} (\mathrm{eV}/c)^{-1}$.

Energy- or temperature-like quantities can most conveniently be measured in
terms of the Fermi energy or Fermi temperature which are equivalent in our
units,
\begin{equation}
\epsilon_{\text{F}}\equiv T_{\text{F}}= \frac{\kF^2}{2M} \equiv \kF^2. 
\end{equation}
As $1\equiv 2M = 2 \frac{M}{\mathrm{eV}/c^2}\, \mathrm{eV}/c^2$, fixing the
atom mass $M$ in units of eV/$c^2$ (e.g., $M_{{}^6\mathrm{Li}}\simeq 5.6\times
10^{9} \mathrm{eV}/c^2$ for Lithium-6) corresponds to expressing the velocity of light $c$
in units of $\sqrt{\mathrm{eV}}$. This establishes eV units for our
energy-like quantities such as the Fermi energy.  

As shown in the main text, the scattering length $a$ remains as the only
physical parameter of the interacting Fermi gas in the broad-resonance
limit. Together with the temperature, the relevant dimensionless quantities
characterizing the phase diagram are
\begin{equation}
c^{-1}=(a \kF)^{-1}, \quad \frac{T}{T_{\text{F}}} = \frac{T}{\kF^2}, 
\end{equation}
where $c$ (not to be confused with the velocity of light here) can be viewed
as a concentration parameter, as it measures the inverse scattering length in
units of the interparticle spacing $\sim \kF^{-1}$.

We use our result for the critical temperature at unitarity,
$T_{\text{c}}/T_{\text{F}}\simeq 0.276$ at $a \kF=0$, to illustrate how to convert our
results into standard laboratory units, using Lithium-6 as an example. First,
we assume an atom density of $n=10^{13} \mathrm{cm}^{-3}$ which implies
$\kF\simeq 1.3\mathrm{eV}/c$. The resulting critical temperature reads
$T_{\text{c}}\simeq 0.276 T_{\text{F}}\simeq0.477 (\mathrm{eV}/c)^2$. In our
units, where the Lithium-6 mass obeys $2M=1$, the velocity of light can be
expressed as $c= \sqrt{2 \frac{M}{\mathrm{eV}/c^2} \mathrm{eV}}\simeq 1.06
\times 10^5 \sqrt{\mathrm{eV}}$. Inserting this into the critical temperature
yields $T_{\text{c}}\simeq 4.26\times 10^{-11}\mathrm{eV}\simeq 0.49
\mu\mathrm{K}$.

\section{Symmetries}
\label{sec:symmetries}

Symmetries of the microscopic action $S$ are also symmetries of the
effective action $\Gamma$, provided the functional measure does not
give rise to an anomaly and that the infrared cutoff respects the
symmetry. The symmetries of $\Gamma$ severely restrict its possible
form. They are therefore helpful in devising useful truncations. All
truncations should be consistent with the symmetries.

In our case we have first the symmetries of translations in space and
time and the rotation symmetry in space. Further, a continuous U(1)
symetry reflects the conserved number of atoms.
\begin{eqnarray}
  \nonumber
  \psi_1\rightarrow e^{i\alpha}\psi_1, &\quad& \psi_1^\ast\rightarrow 
  e^{-i\alpha}\psi_1^\ast\\
  \nonumber
  \psi_2\rightarrow e^{i\alpha}\psi_2, &\quad& \psi_2^\ast\rightarrow 
  e^{-i\alpha}\psi_2^\ast\\
  \varphi\rightarrow e^{2i\alpha}\varphi, &\quad& \varphi^\ast
  \rightarrow e^{-2i\alpha}\varphi^\ast.
\end{eqnarray}
This symmetry is conserved by the cutoff and easily realized in
truncations by including in $\Gamma$ only terms with total atom number
zero. (Here $\psi$ and $\varphi$ have atom number one and two, and
$\psi^*$, $\varphi^*$ the corresponding negative numbers.) The action
\eqref{eq:action} also shows an SU(2)-symmetry of ``spin rotations''
between the two fermion components. This is again easily
implemented. Further discrete symmetries are parity reflections
($\psi(\tau,\vec x)\to\psi(\tau,-\vec x)$, $\varphi(\tau,\vec
x)\to\varphi(\tau,-\vec x)$) as well as euclidean time reflection
\begin{eqnarray}
\nonumber
\psi_1(\tau,\vec{x}) \rightarrow & \psi_1^\ast(-\tau,\vec{x}),  \quad &
\psi_1^\ast(\tau,\vec{x}) \rightarrow -\psi_1(-\tau,\vec{x}),\\
\nonumber
\psi_2(\tau,\vec{x}) \rightarrow & -\psi_2^\ast(-\tau,\vec{x}), \quad &
\psi_2^\ast(t,\vec{x}) \rightarrow \psi_2(-\tau,\vec{x}),\\
\varphi(\tau,\vec{x}) \rightarrow & \varphi^\ast(-\tau,\vec{x}), \quad &
\varphi^\ast(\tau,\vec{x}) \rightarrow \varphi(-\tau,\vec{x}). 
\label{eq:timereversalsymm}
\end{eqnarray}
We note that $\psi_\alpha$ and $\psi_\alpha^\ast$ are separate
Grassmann variables for which the structure of complex conjugation is
not defined a priori. The transformation \eqref{eq:timereversalsymm}
is actually not compatible with the usual complex structure. (The
action \eqref{eq:action} is also invariant under $\varphi(\tau,\vec
x)\leftrightarrow \varphi^*(-\tau,\vec x)$, $\psi_\alpha(\tau,\vec
x)\leftrightarrow \psi_\alpha(-\tau,\vec x)$ if all Grassmann
variables are completely reordered.) Time reflection symmetry does not
forbid terms with an even number of $\tau$-derivatives.

For zero temperature no rest frame is singled out a priori and we
expect invariance under Galilei-transformations. The implications of
Galilean invariance can be understood best if we perform an analytic
continuation from euclidean time to real time $\tau=it$. The
microscopic action \eqref{eq:action} becomes then
\begin{eqnarray}
  \nonumber 
  &&\hspace{-1.2cm} 
  S[\varphi,\psi]=i\int_{-\infty}^{\infty} d t \int d^3 x 
  \Bigl(\psi^\dagger \left(-i\partial_t 
    -\Delta- \mu \right)\psi\\\nonumber 
  &&\hspace{1.2cm}  +\varphi^\ast  \left(-i\partial_t 
    -\frac{1}{2}\Delta+\nu - 2 \mu \right)\varphi\\
  &&\hspace{1.2cm}- h_\varphi 
  \left(\varphi^\ast \psi^T\epsilon \psi-\varphi \psi^\dagger \epsilon \psi^\ast
  \right)\Bigr)\,.
\label{eq:microscopicrealtimeaction}
\end{eqnarray}
Galilean boost transformations (with a boost velocity
$2\vec{q}$) act on the fields as
\begin{eqnarray}
  \nonumber
  \psi(t,\vec{x})\rightarrow \psi^\prime(t,\vec{x}) &=&
  e^{-i(\vec{q}^2t-\vec{q}\vec{x})}\psi(t,\vec{x}-2\vec{q}t)\\
  \varphi(t,\vec{x})\rightarrow 
  \varphi^\prime(t,\vec{x}) &=& e^{-2i(\vec{q}^2t-
    \vec{q}\vec{x})}\varphi(t,\vec{x}-2\vec{q}t).
\end{eqnarray}
While the invariance of the Yukawa term under this transformation is
obvious, its realization for the kinetic term is more involved.
Performing the transformation explicitly one finds
\begin{eqnarray}
  \nonumber
  \psi^\dagger\Delta\psi &\rightarrow& \psi^\dagger \Delta 
  \psi-\vec{q}^2\psi^\dagger \psi+2i\vec{q}\psi^\dagger \vec{\nabla}\psi\\ 
  \nonumber
  \varphi^\dagger\Delta\varphi &\rightarrow& \varphi^\dagger 
  \Delta \varphi-4\vec{q}^2\psi^\dagger \varphi+4i\vec{q}
  \varphi^\dagger \vec{\nabla}\varphi\\
  \nonumber
  \psi^\dagger i\partial_t \psi & \rightarrow & \psi^\dagger 
  i\partial_t \psi + \vec{q}^2\psi^\dagger \psi-2i \vec{q}
  \psi^\dagger \vec{\nabla}\psi\\ 
  \varphi^\dagger i\partial_t \varphi & \rightarrow & 
  \varphi^\dagger i\partial_t \varphi + 2\vec{q}^2
  \varphi^\dagger \varphi-2i \vec{q}\varphi^\dagger \vec{\nabla}\varphi,
\end{eqnarray}
such that indeed the combinations 
\begin{equation}
-i\partial_t-\Delta
\label{eq:Galileaninvderivativfer}
\end{equation} 
for the fermions and
\begin{equation}
-i\partial_t-\frac{\Delta}{2}
\label{eq:Galileaninvderivativbos}
\end{equation} 
for the bosons lead to this invariance. The conserved Noether charge
of Galilean symmetry is the center of mass momentum.

In the case of a bosonic condensate $\varphi(\vec x, t)=\varphi_0$ the
Galilean symmetry is spontaneously broken. Now the condensate
determines the rest frame. The spontaneous symmetry breaking of
Galilean symmetry comes in pair with spontaneous symmetry breaking of
the U(1) symmetry associated to atom number. It can be associated with
superfluidity. We stress that for $T>0$ Galilean symmetry is
explicitly broken since the heat bath singles out a rest frame. For
low temperatures truncations consistent with Galilean transformations
(and therefore always containing the combinations
\eqref{eq:Galileaninvderivativfer} and
\eqref{eq:Galileaninvderivativbos} remain useful, however, since
symmetry breaking terms vanish for $T\to0$.

Finally, we observe another useful invariance of the action for
$T=0$. It arises if we consider $\mu$ as a source term for the fermion
and boson bilinears and consider transformations that also change
$\mu$. (These type of invariances are not symmetries of the action in
the standard sense, since ``parameters'' of the action are changed by
the transformation.) Let us extend
Eq. \eqref{eq:microscopicrealtimeaction} to a $t$-dependent source
$\mu(t)$. There is now a semilocal $U(1)$ invariance of the form
\begin{eqnarray}
  \nonumber
  \psi\rightarrow e^{i\alpha(t)}\psi, 
  \quad \psi^\dagger\rightarrow e^{-i\alpha(t)}\psi^\dagger\\
  \nonumber
  \varphi\rightarrow e^{2i\alpha(t)}\varphi, 
  \quad \varphi^\ast \rightarrow e^{-2i\alpha(t)}\varphi^\ast\\
  \mu\rightarrow\mu+\partial_t \alpha.
\end{eqnarray}
This holds, since the combinations $(-i\partial_t-\mu)$ and
$(-i\partial_t-2\mu)$ act as covariant derivatives for the fermions
and bosons, respectively.

Combining semi-local $U(1)$ symmetry and Galilean symmetry at $T=0$ we
find, that the derivative operators and the chemical potential are
combined to the operator
\begin{equation}
D_\psi=(-i\partial_t-\Delta-\mu)
\end{equation}
for the fermions and
\begin{equation}
D_\varphi=(-i\partial_t-\frac{\Delta}{2}-2\mu)
\end{equation}
for the bosons. In addition to powers of this operator, only spatial
derivative of terms that are invariant under $U(1)$ transformations
like e.g. $\rho \Delta \rho$ with $\rho=\varphi^\ast\varphi$ may
appear.  The above symmetry transformations act linearly on the fields
and so also the effective action $\Gamma$ is invariant. Also there,
only powers of the operators $D_\psi$ and $D_\varphi$ may act on the
fields $\psi$ and $\varphi$.

In the presence of spontaneous symmetry breaking by a nonvanishing
order parameter $\rho=\varphi^\ast\varphi>0$ the possibilities for
combining space and time derivatives get more complex. Effectively,
terms quadratic in $\partial_t$ need not be matched by terms quadratic
in $\Delta$ or $\mu$. For a more detailed discussion see
\cite{FloerchingerWetterich}. Let us finally mention, that there is an
additional conformal invariance at the unitary point in the crossover
where the fermion scattering length diverges $c=a \kF=\infty$
\cite{Son:2005rv}. The only length scale is then set by the chemical
potential $\mu$ or by the particle number density $n$. This gives
further constraints to the effective action at that point in the phase
diagram.

\section{Cutoff functions}
\label{sec:Cutoff}

The computations in the present work are done with regulators which
are optimised w.r.t.\ the derivative expansion, see
e.g. \cite{Litim:2000ci,Litim:2001up,Pawlowski:2005xe}. Such a choice
stabilises the truncation and minimizes the systematical error for the
computed observables. For the scalar field we choose the Litim
regulator \cite{Litim:2000ci}
\begin{eqnarray}\nonumber 
  R_k^{(\varphi)}(q^2)&=&  A_\varphi  k^2 r_{k\varphi}(y)\,, \quad 
  y=\0{q^2}{2k^2}\,,\\[1ex]
  r_{k\varphi}(y)&=&(1-y)\theta(1-y)\,.
\label{eq:regvarphi}\end{eqnarray}
Combining the $\vec q^2$-dependent term in the inverse propagator
$G_\varphi^{-1}$ with the cutoff term one finds for $P_\varphi=\vec
q^2/2 + R_k^{(\varphi)}/A_\varphi$ the simple behavior $P_\varphi =
\vec q^2/2$ for $\vec q^2>2k^2$, $P_\varphi=k^2$ for $\vec
q^2<2k^2$. In other words, our regulator leaves the propagation of
modes with spatial momenta bigger than the cut-off scale, $q^2\geq 2
k^2$, unchanged. For modes with spatial momenta smaller than the
cut-off scale, $q^2<2k^2$, it effectively introduces an infrared mass
term $k^2$.

\begin{figure}[ht]
\centering
\includegraphics[width=0.45\textwidth]{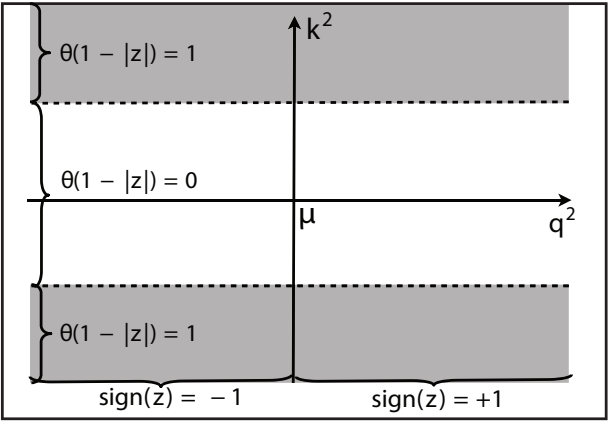}
\caption{The fermionic regulator in the presence of a Fermi surface $\mu>0$.}
\label{fig:FermionicRegulator}
\end{figure}

For the fermionic field we choose an analogue of the scalar regulator
\eq{eq:regvarphi}. Here, however, the propagation of modes close to
the Fermi surface is suppressed,
\begin{eqnarray}\nonumber 
R_k^{(\psi)}(q^2)&=& 
  k^2 r_{k\psi}(z)\,,\\[1ex]
  r_{k\psi}(z)&=&(\text{sign}(z)-z)\theta(1-|z|)\,, 
 \label{eq:regpsi} \end{eqnarray}
where 
\begin{equation} 
z=(q^2-\mu)/k^2.
\end{equation}
This regulator leaves the propagation of modes with spatial momenta
farther away from the Fermi surface than the cut-off scale,
$q^2-\mu\geq k^2$, unchanged but leads to a flat (inverse) propagator
$k^2$ for modes with spatial momenta closer to the Fermi surface than
the cut-off scale $q^2-\mu< k^2$. This effectively introduces an
infrared mass term proportional to $k^2$.  In summary this takes into
account the fact that fermionic fluctuations have to be cut-off
relative to the Fermi surface, and hence introduces to leading order
the same cut-off scales for the fermionic and bosonic fluctuations. The fermionic regulator in the presence of a Fermi surface is displayed in Fig. \ref{fig:FermionicRegulator}.

\section{Flow of the effective potential}
\label{sec:EffetivePotential}

In this App. we derive the flow equation for the effective potential
$U(\rho,\mu)$. We start from the exact flow equation for the average
action (or flowing action) \cite{Wetterich:1992yh}
\begin{equation}
\partial_k \Gamma_k =\frac{1}{2} \text{STr} (\Gamma_k^{(2)}+R_k)^{-1}\partial_k R_k.
\end{equation}
This is evaluated for a bosonic field $\bar\varphi$ that is constant
in space and in (Matsubara-) time, $\bar\varphi(\vec
x,\tau)=\bar\varphi$. (The expectation value of the fermionic field
$\psi$ vanishes -- for vanishing fermion source -- due to its
Grassmann property.) Using the truncation in Eq. \eqref{2H} we find
the flow equation for the potential
\begin{eqnarray}
\nonumber
\partial_k U_k{\big |}_{\bar\rho} &=& \frac{1}{2} \int\limits_{q_0,\vec q} {\bigg \{} 
\left[(G_{k\varphi})_{11}+(G_{k\varphi})_{22}\right]A_\varphi^{-1}\partial_k \left(A_\varphi k^2 r_{k\varphi}\right)\\
\nonumber
&-& \left[ (G_{k\psi})_{13}+(G_{k\psi})_{24}-(G_{k\psi})_{31}-(G_{k\psi})_{42}\right]\\
&& \times \partial_k\left(k^2 r_{k\psi}\right) {\bigg \} },
\label{eq:flowUk1}
\end{eqnarray}
which is evaluated for fixed $\bar \rho=\bar
\varphi^*\bar\varphi$. The dimensionless function $r_{k\varphi}$
depends on $y=\vec q^2/(2k^2)$ while $r_{k\psi}$ depends on $z=(\vec
q^2-\mu_0)/k^2$. The regularized propagators $G_{k\varphi}$ and
$G_{k\psi}$ that appear in Eq. \eqref{eq:flowUk1} are modified from
Eqs.\ \eqref{eq:1XY} \eqref{2I} by the presence of the ultraviolet
regulator
\begin{eqnarray}
\nonumber
G_{k\varphi}^{-1} &=& G_\varphi^{-1}+\begin{pmatrix}  k^2 r_{k\varphi} && 0 \\ 0 && k^2 r_{k\varphi} \end{pmatrix}\\
G_{k\psi}^{-1}  &=& G_\psi^{-1} + \begin{pmatrix} 0 && -k^2 r_{k\psi} \\ k^2 r_{k\psi} && 0 \end{pmatrix}.
\end{eqnarray}
We can now perform the summation over the Matsubara frequencies
$q_0=2\pi T n$ for the bosons and $q_0=2\pi T (n+1/2)$ for the
fermions. The integration over $\vec q$ is performed quite generally
in $d$ spatial dimensions.  The result can be expressed in terms of
the dimensionless variables
\begin{eqnarray}\label{DimlessVar}
  \nonumber
  w_1=\frac{U_k^\prime}{k^2},\quad w_2=\frac{U_k^\prime+2\rho U_k^{\prime\prime}}{k^2},\quad w_3=\frac{h_\varphi^2 \rho}{k^4},\\
  \tilde \mu=\frac{\mu_0}{k^2},\quad \Delta\tilde\mu=\frac{\mu-\mu_0}{k^2},\quad\tilde T=\frac{T}{k^2},\quad S_\varphi=\frac{Z_\varphi}{A_\varphi},
\end{eqnarray}
and the anomalous dimension 
\begin{equation}\label{AnDim}
\eta_{A_\varphi}=-\frac{k\partial_k A_\varphi}{A_\varphi}.
\end{equation}
A term involving the anomalous dimension appears if we evaluate the
flow for fixed $\rho=A_\varphi\bar\rho$ instead of fixed $\bar
\rho$. The flow of the effective potential reads now
\begin{eqnarray}
\nonumber
k\partial_k U_k &=& \eta_{A_\varphi} \,\rho\, U_k^\prime+8\sqrt{2} \frac{k^{d+2} v_d }{d\,S_\varphi} \left(1-\frac{2}{d+2}\eta_{A_\varphi}\right) s_{\text{B}}^{(0)}\\
&& - \,8  \frac{k^{d+2} v_d}{d}\,l(\tilde \mu)\, s_{\text{F}}^{(0)}.
\label{eq:flowofUk2}
\end{eqnarray}
Here, the coefficient $v_d$ is proportional to the surface of the
$d$-dimensional unit sphere, which is $(2\pi)^d 4 v_d$, with
$v_d^{-1}=2^{d+1}\pi^{d/2} \Gamma(d/2)$. In particular one has
$v_3=1/(8\pi^2)$.

The threshold functions
$s_{\text{B}}^{(0)}=s_{\text{B}}^{(0)}(w_1,w_2,\tilde
T,S_\varphi,\eta_{A_\varphi})$ and
$s_{\text{F}}^{(0)}=s_{\text{F}}^{(0)}(w_3,\tilde
\mu,\Delta\tilde\mu,\tilde T)$ as well as the function $l(\tilde \mu)$
used in Eq. \eqref{eq:flowofUk2} depend on the choice of the infrared
regulator functions $r_{k\varphi}$ and $r_{k\psi}$. They describe the
decoupling of modes when the effective ``masses'' $w_j$ or $-\tilde
\mu$ get large. The threshold functions for the bosonic fluctuations
reads
\begin{gather}
\nonumber
\left(1-\frac{2\,\eta_{A_\varphi}}{d+2}\right)s_{\text{B}}^{(0)} = d\int\limits_0^\infty dy\, y^{\frac{d}{2}-1} \left(r_{k\varphi}-y r_{k\varphi}^\prime-\eta_{A_\varphi}r_{k\varphi}\right)\\
\nonumber
\times \frac{\frac{1}{2}(w_1+w_2)+y+r_{k\varphi}}{\sqrt{w_1+y+r_{k\varphi}}\sqrt{w_2+y+r_{k\varphi}}}\\
\nonumber
\times \left[\frac{1}{2}+N_{\text{B}}(\sqrt{w_1+y+r_{r\varphi}}\sqrt{w_2+y+r_{k\varphi}}/S_\varphi)\right],\\
\end{gather}
and involves the Bose function
\begin{equation}\label{BoseF}
N_{\text{B}}(\epsilon)=\frac{1}{e^{\epsilon/\tilde T}-1}.
\end{equation}
and the optimized cutoff function \cite{Litim:2000ci}
\begin{equation}
r_{k\varphi}(y)=(1-y)\theta(1-y).
\end{equation}
We normalize the threshold function such that $s_{\text{B}}^{(0)}=1$ for $w_1=w_2=\tilde T=0$. For our cutoff one finds the particular simple expression
\begin{eqnarray}
\nonumber
s_{\text{B}}^{(0)}&=&\left[\sqrt{\frac{1+w_1}{1+w_2}}+\sqrt{\frac{1+w_2}{1+w_1}}\right]\\
&&\times\left[\frac{1}{2}+N_{\text{B}}(\sqrt{1+w_1}\sqrt{1+w_2}/S_\varphi)\right].
\label{eq:Bosonicth}
\end{eqnarray}

The threshold function for the fermionic fluctuations is obtained similar. For a generic cutoff that addresses the spatial momentum, it reads
\begin{eqnarray}
\nonumber
l(\tilde \mu) s_{\text{F}}^{(0)}=d\int_{-\tilde\mu}^\infty dz\,(z+\tilde\mu)^{\frac{d}{2}-1} \left(r_{k\psi}-z r_{k\psi}^\prime\right)\\
\nonumber
\times \frac{(z+r_{k\psi}-\Delta\tilde\mu)}{\sqrt{w_3+(z+r_{k\psi}-\Delta\tilde\mu)^2}}\\
\times \left[\frac{1}{2}-\,N_{\text{F}}\left(\sqrt{w_3+(z+r_{k\psi}-\Delta\tilde \mu)^2}\right)\right].
\label{eq:thresholdfermionsimplicitdef}
\end{eqnarray}
Here we employ the Fermi function
\begin{equation}\label{FermiF}
N_{\text{F}}(\epsilon)=\frac{1}{e^{\epsilon/\tilde T}+1}.
\end{equation}
Note that for a generic cutoff the right hand side of equation \eqref{eq:thresholdfermionsimplicitdef} does not necessarily factorize. In that case one might work with a threshold function $s_{\text{F}}^{(0)}$ that also depends on $\tilde \mu$. For our cutoff
\begin{equation}
r_{k\psi}=(\text{sign}(z)-z)\theta(1-|z|).
\end{equation}
one has $s_{\text{F}}^{(0)}=1$ for $w_3=\tilde\mu=\Delta\tilde\mu=0$. For $\mu=\mu_0$ and therefore $\Delta\tilde\mu=0$ the threshold function assumes the simple form
\begin{equation}
\nonumber
s_{\text{F}}^{(0)}=\frac{2}{\sqrt{1+w_3}}\left[\frac{1}{2}-\,N_{\text{F}}\left(\sqrt{1+w_3}\right)\right]
\label{eq:Fermionicth}
\end{equation}
while
\begin{equation}
l(\tilde \mu)=\theta(\tilde \mu+1)(\tilde \mu+1)^{d/2}-\theta(\tilde \mu-1)(\tilde \mu-1)^{d/2}.
\end{equation}
In the limit $\tilde T=T/k^2\to0$ the thermal contributions to the flow of $U_k$ vanish, $N_{\text{B}}=N_{\text{F}}=0$.

Taking a derivative with respect to $\rho$ on both sides of Eq. \eqref{eq:flowofUk2} we obtain
\begin{eqnarray}
\nonumber
k\partial_k U_k^\prime &=& \eta_{A_\varphi}(U_k^\prime+\rho U_k^{\prime\prime})-8\sqrt{2} \frac{k^d v_d}{dS_\varphi}\left(1-\frac{2}{d+2}\eta_{A_\varphi}\right)\\
\nonumber
&&\times \left[U_k^{\prime\prime}s_{\text{B}}^{(1,0)}+(3U_k^{\prime\prime}+2\rho U_k^{(3)})s_{\text{B}}^{(0,1)}\right]\\
&& +8\frac{k^{d-2}v_d}{d} h_\varphi^2\,l(\tilde \mu)\,s_{\text{F}}^{(1)}.
\end{eqnarray}
Here we introduce the derivatives of the threshold functions
\begin{eqnarray}
\nonumber
s_{\text{B}}^{(1,0)} &=& -\frac{\partial}{\partial w_1}s_{\text{B}}^{(0)}\\
\nonumber
s_{\text{B}}^{(0,1)} &=& -\frac{\partial}{\partial w_2}s_{\text{B}}^{(0)}\\
\nonumber
s_{\text{F}}^{(1)} &=& -\frac{\partial}{\partial w_3}s_{\text{F}}^{(0)}.\\
\end{eqnarray}
We may divide these into contributions from quantum and thermal fluctuations
\begin{eqnarray}
\nonumber
s_{\text{B}}^{(1,0)}&=&(w_2-w_1) s_{\text{B,Q}}^{(1,0)}+s_{\text{B},T}^{(1,0)},\\
\nonumber
s_{\text{B}}^{(0,1)}&=&(w_2-w_1) s_{\text{B,Q}}^{(0,1)}+s_{\text{B},T}^{(0,1)},\\
s_{\text{F}}^{(1)}&=& s_{\text{F,Q}}^{(1)}+s_{\text{F},T}^{(1)}.
\end{eqnarray}
For $\tilde T\to0$ the thermal contribution vanishes $s_{\text{B},T}^{(0,1)}=s_{\text{B},T}^{(1,0)}=s_{\text{F},T}^{(1)}=0$. 
We extracted a factor $(w_2-w_1)$ from the threshold functions $s_{\text{B,Q}}^{(1,0)}$ and $s_{\text{B,Q}}^{(0,1)}$ to make explicit that these contributions vanish for $w_1=w_2$, which holds for $\rho=0$.

For our choice of the regulator functions $r_{k,\psi}$ and $r_{k\varphi}$ we find the explicit expressions
\begin{eqnarray}
\nonumber
s_{\text{B,Q}}^{(1,0)} &=& \frac{1}{4(1+w_1)^{3/2}(1+w_2)^{1/2}},\\
\nonumber
s_{\text{B,Q}}^{(0,1)} &=& -\frac{1}{4(1+w_1)^{1/2}(1+w_2)^{3/2}},\\
\nonumber
s_{\text{F,Q}}^{(1)} &=& \frac{1}{2(1+w_3)^{3/2}}.
\end{eqnarray}
For $\tilde T>0$ the thermal fluctuations lead to the additional contributions from the bosons
\begin{eqnarray}
\nonumber
s_{\text{B},T}^{(1,0)}&=& 2(w_2-w_1)\, s_{\text{B,Q}}^{(1,0)}\, N_{\text{B}}\left(\sqrt{(1+w_1)(1+w_2)}/S_\varphi\right)\\
\nonumber
&+& s_{\text{B,Q}}^{(0)}\frac{\sqrt{1+w_2}}{\sqrt{1+w_1}S_\varphi}\, N_{\text{B}}^\prime\left(\sqrt{(1+w_1)(1+w_2)}/S_\varphi\right),\\
\nonumber
s_{\text{B},T}^{(0,1)}&=& 2(w_2-w_1)\, s_{\text{B,Q}}^{(1,0)}\, N_{\text{B}}\left(\sqrt{(1+w_1)(1+w_2)}/S_\varphi\right)\\
\nonumber
&+& s_{\text{B,Q}}^{(0)}\frac{\sqrt{1+w_1}}{\sqrt{1+w_2}S_\varphi}\, N_{\text{B}}^\prime\left(\sqrt{(1+w_1)(1+w_2)}/S_\varphi\right).\\
\end{eqnarray}
Here, we use the derivative of the Bose function
\begin{equation}
N_{\text{B}}^\prime(\epsilon)=\frac{\partial}{\partial \epsilon}N(\epsilon).
\end{equation}
Similarly, the fermionic part of the thermal contribution reads
\begin{eqnarray}
\nonumber
s_{\text{F},T}^{(1)} &=& -2 \,s_{\text{F,Q}}^{(1)}\,N_{\text{F}}\left(\sqrt{1+w_3}\right)\\
&&-s_{\text{F,Q}}^{(0)}\frac{1}{\sqrt{1+w_3}}\, N_{\text{F}}^\prime\left(\sqrt{1+w_3}\right),
\end{eqnarray}
with the derivative of the Fermi function
\begin{equation}
N_{\text{F}}^\prime(\epsilon)=\frac{\partial}{\partial \epsilon} N_{\text{F}}(\epsilon).
\end{equation}

\section{Gradient Coefficient and Wave Function Renromalization}
\label{app:GradientCoeff}

We compute the flow equations for the kinetic coefficients $A_\varphi$, $Z_\varphi$ for the bosons according to the following procedure: First, we obtain the frequency and momentum dependent one-loop correction to the inverse boson propagator in the presence of the infrared cutoff $R_k$. The flow equations for the boson propagator are then obtained by taking the appropriate cutoff derivative. Finally, we project onto the flow equations of $A_\varphi$ and $Z_\varphi$ by taking derivatives with respect to the momentum $\vec q^2$ and the frequency $q_0$, respectively. This procedure is equivalent with the derivation of the exact flow equation for the inverse propagator \cite{Wetterich:1992yh}, \cite{Berges:2000ew}, and a truncation where only the couplings appearing in Eq.\ \eqref{2H} are taken into account. 

We work in momentum space with a basis of bare real field fluctuations $\delta\bar\varphi_1,\delta\bar\varphi_2$, which relate to the complex bosonic fields in momentum space as
\begin{eqnarray}
\delta \bar \varphi (Q) &=& (\delta\bar\varphi_1 (Q) + i \delta \bar\varphi_2 (-Q))/\sqrt{2},\\\nonumber
\delta \bar\varphi^* (Q) &=& (\delta\bar\varphi_1 (-Q) - i \delta\bar\varphi_2 (Q))/\sqrt{2}.
\end{eqnarray}
The one-loop correction to the inverse boson propagator $\Delta \bar P$ is then defined with
\begin{equation}
 (\Delta\bar P_\varphi)_{ab}(q)\delta(q+q^\prime) = \frac{\delta^2 \Delta \Gamma}{\delta\bar\varphi_a(q)
\delta\bar\varphi_b(q^\prime)}\Big|_{\bar\varphi_1 =
\sqrt{2\bar\rho_0},\bar \varphi_2=0} .  
\end{equation}
Here $\Delta \Gamma$ is the one-loop correction to the effective action. It is evaluated for a constant background field $\bar\varphi_1=\sqrt{2\bar \rho_0}$ if the minimum of $U_k$ occurs for $\bar \rho_0(k)\neq 0$, while $\bar \varphi_1=0$ in the symmetric regime where $\bar \rho_0(k)=0$. Then the wave function renormalization and gradient coefficient at one loop order are obtained as 
\begin{eqnarray}\label{WfrGradC}
\Delta Z_\varphi &=& -
\frac{\partial}{\partial q_0} (\Delta\bar P_\varphi)_{12}(q_0,0){\big |}_{q_0=0} ,\\\nonumber
\Delta A_\varphi &=& 2 \frac{\partial}{
  \partial \vec q\,^2} (\Delta\bar P_\varphi)_{22}(0,\vec q){\big |}_{\vec q=0}.
\end{eqnarray}  

Within our truncation, $(\Delta \bar P_\varphi)_{12}$ and $(\Delta \bar P_\varphi)_{22}$ receive contributions from fermionic and from bosonic fluctuations,
\begin{widetext}
\begin{eqnarray}
(\Delta\bar P_\varphi)_{12}(q_0,0) &=& (\Delta\bar P^{\mathrm{(F)}}_\varphi)_{12}(q_0,0) + (\Delta\bar P^{\mathbf{(B)}}_\varphi)_{12}(q_0,0),\\\nonumber
(\Delta\bar P^{\mathrm{(F)}}_\varphi)_{12}(q_0,0) &=& q_0 \bar h_\varphi^2  \int_p \frac{f(\vec p)}{\det_{\mathrm{F}}(p_0, \vec p) \det_{\mathrm{F}}(p_0+q_0, \vec p) },\\\nonumber
(\Delta\bar P^{\mathbf{(B)}}_\varphi)_{12}(q_0,0) &=&  Z_\varphi q_0 \bar \lambda_\varphi^2\bar \rho_0  \int_p\frac{3 a_{22}(\vec p) - a_{11}(\vec p)}{\det_{\mathrm{B}}(p_0, \vec p) \det_{\mathrm{B}}(p_0+q_0, \vec p ) } ,\\\nonumber
(\Delta\bar P_\varphi)_{22}(0,\vec q) &=& (\Delta\bar P^{\mathrm{(F)}}_\varphi)_{22}(0,\vec q) + (\Delta\bar P^{\mathbf{(B)}}_\varphi)_{22}(0,\vec q),\\\nonumber
(\Delta\bar P^{\mathrm{(F)}}_\varphi)_{22}(0,\vec q) &=&  - \bar h_\varphi^2\int_p \frac{p_0^2 + f(\vec p) f(\vec p + \vec q) + \bar h_\varphi^2\bar \rho_0}{\det_{\mathrm{F}}(p_0, \vec p) \det_{\mathrm{F}}(p_0, \vec p + \vec q) },\\\nonumber
(\Delta\bar P^{\mathbf{(B)}}_\varphi)_{22}(0,\vec q) &=&- \bar \lambda_\varphi^2\bar\rho_0 \int_p\frac{2 ( Z_\varphi p_0)^2 +  a_{11}(\vec p+ \vec q )a_{22}(\vec p )  + a_{11}(\vec p )a_{22}(\vec p + \vec q)}{\det_{\mathrm{B}}(p_0, \vec p) \det_{\mathrm{B}}(p_0, \vec p + \vec q) }.
\label{eq:oneloopcontrbosprop}
\end{eqnarray}
\end{widetext}
Here we use the abbreviations 
\begin{eqnarray}
f(\vec p) &=&   \vec p\,^2 - \mu +R_k^{(\psi)}(\vec p^2),\\\nonumber
\text{det}_F(\omega, \vec p) &=& p_0^2 + f^2(\vec p) +\bar h_\varphi^2\bar \rho_0,\\\nonumber
a_{11}(\vec p ) &=& A_\varphi \vec p\,^2/2 + 2\bar \lambda_\varphi\bar \rho_0 +R_k^{(\varphi)}(\vec p^2),\\\nonumber
a_{22}(\vec p ) &=& A_\varphi \vec p\,^2/2 + +R_k^{(\varphi)}(\vec p^2),\\\nonumber
\text{det}_B(\omega, \vec p) &=& ( Z_\varphi p_0)^2 + a_{11}(\vec p ) a_{22}(\vec p ).
\end{eqnarray}
(We work with the boson propagator parameterized appropriately for the
regime with U(1) symmetry breaking. In the symmetric regime the
contributions from bosonic fluctuations vanish.)  To obtain the flow
equation for the boson propagator $(\bar P_\varphi)_{12}$, we take the
cutoff derivative of the expressions in Eq.\
\eqref{eq:oneloopcontrbosprop}
\begin{widetext}
\begin{eqnarray}
\nonumber
(\partial_k \bar P_\varphi)_{12}(q_0,0) &=& (\partial_k \bar P_\varphi^{\mathrm{(F)}})_{12}(q_0,0)+(\partial_k \bar P_\varphi^{\mathbf{(B)}})_{12}(q_0,0),\\
\nonumber
(\partial_k \bar P_\varphi^{\mathrm{(F)}})_{12}(q_0,0) &=& \tilde \partial_k (\Delta \bar P_\varphi^{\mathrm{(F)}})_{12}(q_0,0)\\
\nonumber
&=& q_0 \bar h_\varphi^2 \int_p {\bigg \{} \frac{\partial_k R_k^{(\psi)}(\vec p^2)}{\text{det}_F(p_0,\vec p)\, \text{det}_F(p_0+q_0,\vec p)}\\
\nonumber
&&
- \frac{2f(\vec p)^2 \, \partial_k R_k^{(\psi)}(\vec p^2)}{\text{det}_F^2(p_0,\vec p)\, \text{det}_F(p_0+q_0,\vec p)}
- \frac{2f(\vec p)^2 \, \partial_k R_k^{(\psi)}(\vec p^2)}{\text{det}_F(p_0,\vec p)\, \text{det}_F^2(p_0+q_0,\vec p)}{\bigg \}},\\
\nonumber
(\partial_k \bar P_\varphi^{\mathbf{(B)}})_{12}(q_0,0) &=& \tilde \partial_k (\Delta \bar P_\varphi^{\mathbf{(B)}})_{12}(q_0,0)\\
\nonumber
&=& Z_\varphi q_0 \bar \lambda_\varphi^2 \bar \rho_0 \int_p {\bigg \{} \frac{2 \partial_k R_k^{(\varphi)}(\vec p^2)}{\text{det}_B(p_0,\vec p)\, \text{det}_B(p_0+q_0,\vec p)}\\
\nonumber
&&- \frac{[3a_{22}(\vec p)-a_{11}(\vec p)][a_{11}(\vec p)+a_{22}(\vec p)]\partial_k R_k^{(\varphi)}(\vec p^2)}{\text{det}_B^2(p_0,\vec p)\,\text{det}_B(p_0+q_0,\vec p)}\\
&&- \frac{[3a_{22}(\vec p)-a_{11}(\vec p)][a_{11}(\vec p)+a_{22}(\vec p)]\partial_k R_k^{(\varphi)}(\vec p^2)}{\text{det}_B(p_0,\vec p)\,\text{det}_B^2(p_0+q_0,\vec p)}{\bigg \}}.
\label{eq:flowP12}
\end{eqnarray}
Here the symbol $\tilde \partial_k$ denotes a formal derivative that
hits only the cutoff term $R_k$. Similarly, we obtain the flow
equation for $(\bar P_\varphi)_{22}$ as a function of the spatial
momentum $\vec q$
\begin{eqnarray}
\nonumber
(\partial_k \bar P_\varphi)_{22}(0,\vec q) &=& (\partial_k \bar P_\varphi^{\mathrm{(F)}})_{22}(0,\vec q) + (\partial_k \bar P_\varphi^{\mathbf{(B)}})_{22}(0,\vec q),\\
\nonumber
(\partial_k \bar P_\varphi^{\mathrm{(F)}})_{22}(0,\vec q) &=& \tilde \partial_k (\Delta \bar P_\varphi^{\mathrm{(F)}})_{22}(0,\vec p)\\
\nonumber
&=& -\bar h_\varphi^2 \int_p {\bigg \{} \frac{f(\vec p+\vec q) \partial_k R_k^{(\psi)}(\vec p^2)}{\text{det}_F(p_0,\vec p)\,\text{det}_F(p_0,\vec p+\vec q)}\\
\nonumber
&& - \frac{2[p_0^2+f(\vec p)f(\vec p+\vec q)+\bar h_\varphi^2 \bar \rho_0] f(\vec p) \partial_k R_k^{(\psi)}(\vec p^2)}{\text{det}_F^2(p_0,\vec p)\,\text{det}_F(p_0,\vec p+\vec q)}\\
\nonumber
&&+\frac{f(\vec p-\vec q) \partial_k R_k^{(\psi)}(\vec p^2)}{\text{det}_F(p_0,\vec p-\vec q)\,\text{det}_F(p_0,\vec p)}\\
\nonumber
&& - \frac{2[p_0^2+f(\vec p-\vec q)f(\vec p)+\bar h_\varphi^2 \bar \rho_0] f(\vec p) \partial_k R_k^{(\psi)}(\vec p^2)}{\text{det}_F(p_0,\vec p-\vec q)\,\text{det}_F^2(p_0,\vec p)}{\bigg \}},\\
\nonumber
(\partial_k \bar P_\varphi^{\mathbf{(B)}})_{22}(0,\vec q) &=& \tilde \partial_k (\Delta \bar P_\varphi^{\mathbf{(B)}})_{22}(0,\vec q)\\
\nonumber
&=& -\bar \lambda_\varphi^2 \bar \rho_0 \int_p {\bigg \{} \frac{[a_11(\vec p+\vec q)+a_{22}(\vec p+\vec q)]\partial_k R_k^{(\varphi)}(\vec p^2)}{\text{det}_B(p_0,\vec p)\,\text{det}_B(p_0,\vec p+\vec q)}\\
\nonumber
&&- \frac{[2(Z_\varphi p_0)^2+a_{11}(\vec p+\vec q) a_{22}(\vec p)+a_{11}(\vec p) a_{22}(\vec p+\vec q)][a_{11}(\vec p)+a_{22}(\vec p)]\partial_k R_k^{(\varphi)}(\vec p^2)}{\text{det}_B^2(p_0,\vec p)\,\text{det}_B(p_0,\vec p+\vec q)}\\
\nonumber
&&+\frac{[a_11(\vec p-\vec q)+a_{22}(\vec p-\vec q)]\partial_k R_k^{(\varphi)}(\vec p^2)}{\text{det}_B(p_0,\vec p-\vec q)\,\text{det}_B(p_0,\vec p)}\\
&&- \frac{[2(Z_\varphi p_0)^2+a_{11}(\vec p) a_{22}(\vec p-\vec q)+a_{11}(\vec p-\vec q) a_{22}(\vec p)][a_{11}(\vec p)+a_{22}(\vec p)]\partial_k R_k^{(\varphi)}(\vec p^2)}{\text{det}_B(p_0,\vec p-\vec q)\,\text{det}_B^2(p_0,\vec p)}{\bigg \}}.
\label{eq:flowP22}
\end{eqnarray}
\end{widetext}
In Eq.\ \eqref{eq:flowP22} we performed a shift in the integration variable $\vec p \to \vec p-\vec q$ at several places. From Eqs.\ \eqref{eq:flowP12}, \eqref{eq:flowP22} one obtains the flow equation for the kinetic coeffcients $Z_\varphi$ and $A_\varphi$ by taking approriate derivatives
\begin{eqnarray}
\nonumber
\partial_k Z_\varphi &=& \frac{\partial}{\partial q_0} (\partial_k \bar P_\varphi)_{12}(q_0,0){\big |}_{q_0=0},\\
\partial_k A_\varphi &=& \frac{\partial^2}{\partial q^2} (\partial_k \bar P_\varphi)_{22}(0,\vec q=(q,0,0)){\big |}_{\vec q=0}.
\end{eqnarray}
After this projection and using the cutoff functions in Eqs.\ \eqref{eq:regvarphi}, \eqref{eq:regpsi} we can perfom the Matsubara summation over the frequency $p_0$ and the integration over the momentum  $\vec p$. 
We also switch from the bare couplings to the renormalized ones according to the discussion below Eq. \eqref{2H}; the latter are obtained from rescaling with the flowing gradient coefficient $ A_{\varphi}$. The effect of the gradient coefficient is then encoded in the anomalous dimension Eq.\ \eqref{AnDim}. 
We display the flow equations in the symmetry broken phase, and separate into the fermionic and the bosonic contribution,
\begin{widetext}
\begin{eqnarray}
\nonumber
\eta_{A_\varphi}  &=& \eta^{\mathrm{(F)}}_{A_\varphi} + \eta^{\mathbf{(B)}}_{A_\varphi},\\\nonumber
\eta^{\mathrm{(F)}}_{A_\varphi}&=& \frac{h_\varphi^2 }{6\pi^2 k \sqrt{1+w_3}^{3/2}} \left[ (\tilde\mu +1 )^{3/2} \theta (\tilde\mu +1 ) + (\tilde\mu -1 )^{3/2} \theta (\tilde\mu -1 )\right] \Big(1-2 N_{\mathrm{F}}(\sqrt{1+w_3})+2 \sqrt{1+w_3} N_{\mathrm{F}} '( \sqrt{1+w_3})\Big),\\
\eta^{\mathbf{(B)}}_{A_\varphi} &=&   \frac{\sqrt{2}  \lambda _{\phi }^2 \rho _0}{3 \pi ^2 k  (1+w_2)^{3/2} S_{\varphi }^2} \, \big([1+2 N_{\mathrm{B}}(\sqrt{1+w_2}/S_{\varphi })] S_{\varphi }-2 \sqrt{1+w_2} N_{\mathrm{B}}' (\sqrt{1+w_2}/S_{\varphi })\big).
\end{eqnarray}
The Bose and Fermi functions $N_{\mathrm{B}}, N_{\mathrm{F}}$ are defined in Eqs. \eqref{BoseF}, \eqref{FermiF}, and the derivative refers to the full argument. The other abbreviations are defined in Eq.\ \eqref{DimlessVar}. The flow equation for the rescaled wave function renormalization then reads
\begin{eqnarray}
\nonumber
\partial_t S_\varphi &=& \beta_S + \eta_{A_\varphi} S_\varphi,\quad \beta_S=\beta_S^{\mathrm{(F)}}+\beta_S^{\mathbf{(B)}},\\\nonumber
\beta_S^{\mathrm{(F)}} &=& \frac{h_\varphi^2 }{12\pi^2 k (1+w_3)^{5/2}} \left(2 \tilde\mu ^{3/2} \theta (\tilde\mu )-(\tilde \mu -1 )^{3/2} \theta (\tilde\mu  -1) - (\tilde\mu +1 )^{3/2} \theta (\tilde\mu +1)\right) \\\nonumber
&&\times \big(2 N_{\mathrm{F}}(\sqrt{1+w_3}) (w_3-2)+2  (1+2 \sqrt{1+w_3} N_{\mathrm{F}}' (\sqrt{1+w_3}))\\\nonumber
&&\qquad+2 w_3^2 N_{\mathrm{F}}'' (\sqrt{1+w_3})- w_3 (1+2 \sqrt{1+w_3} N_{\mathrm{F}}' (\sqrt{1+w_3})-2  N_{\mathrm{F}}''(\sqrt{1+w_3}))\big),\\\nonumber
\beta_S^{\mathbf{(B)}}  &=& \frac{\sqrt{2} \lambda_\varphi^2\rho_0 }{3\pi^2k(1 + w_2)^{5/2}S_{\varphi }^2} \big([(-1-2 N_{\mathrm{B}}(\sqrt{1+w_2}/S_{\varphi })) S_{\varphi }^2 + 2 \sqrt{1+w_2} S_{\varphi }N_{\mathrm{B}}'(\sqrt{1+w_2}/S_{\varphi }) ] (2 -w_2-3/4 w_2^2)\\
&&
+w_2 (2 +5/2 w_2+w_2^2/2) N_{\mathrm{B}}'' (\sqrt{1+w_2}/S_{\varphi })\big) .
\end{eqnarray}
\end{widetext}

As mentioned above, the bosonic contributions to the above flow equations are $\sim \rho_0$, thus vanishing in the symmetric phase. The fermionic contributions in  the latter case are obtained via the replacement $\rho_0\to0$, implying $w_3\to 0$. In the zero temperature limit $T/k^2\to 0$ the thermal contributions to the flow equations vanish, formally obtained via the replacement $N_{\mathrm{B}} = N_{\mathrm{F}}=0$.

\section{Flow of the Yukawa coupling}
\label{sec:FlowYukawaCoupling}

In our truncation Eq. \eqref{eq:renormalizedtruncation} the flow of
the Yukawa coupling is quite simple. In the symmetric regime with
$\rho_0=0$ there is no loop contribution such that the flow is given
by the anomalous dimension only,
\begin{equation}
k\partial_k \bar h_\varphi=0, \quad k \partial_k h_\varphi = \frac{1}{2} \eta_{A_\varphi} h_\varphi.
\end{equation}
For a nonvanishing order parameter $\rho_0>0$ there is a loop
contribution $\sim h_\varphi^3 \lambda_\varphi \rho_0$ from a diagram
involving both fermions and bosons. This contribution is expected to
be subleading and we checked numerically that this is indeed the
case. We have omitted this contribution for our numerical result.

The situation is different for an extended truncation where also the effect of
particle hole fluctuations is taken into account. This was described briefly
in Sect.\ \ref{sec:phfluct} and in more detail in \cite{FSDW08}. The
box-diagram Fig. \ref{fig:boxes} gives a contribution to the flow of the
four-fermion interaction $\lambda_\psi$ even though $\lambda_\psi=0$ was
realized on the initial scale $k=\Lambda$ by virtue of a Hubbard-Stratonovich
transformation. On the microscopic scale the only contribution to the
interaction between fermions is given by the exchange of a bound state
$\varphi$, i.~e. a term as in Eq. \eqref{eq:lambdapsitree}. Using the
formalism of rebosonization \cite{Gies:2001nw} one can maintain this picture
also for $k<\Lambda$ and adapt the flow equation for $h_\varphi$ (and possibly
$m_\varphi^2$) such that the effect of the box-diagram in Fig. \ref{fig:boxes}
(the particle-hole contribution) is taken into account. This setting where the
regeneration of the four-fermion coupling $\lambda_\psi$ is suppressed in
favor of a modified flow equation for the Yukawa coupling $h_\varphi$ is
descibed in more detail in \cite{FSDW08}.

\end{document}